\documentclass[useAMS, usenatbib]{mn2e}
\usepackage{graphicx, rotate}
\usepackage{natbib, lscape}
\usepackage{latexsym, amssymb, verbatim}
\usepackage{eufrak}
\usepackage{amsmath}
\usepackage{epsfig}
\usepackage{times}
\usepackage{rotating}
\usepackage{graphicx}
\usepackage{color}
\usepackage{enumerate}

\title[A PDR study of NGC 4038]
{A Photodissociation Region study of NGC 4038}
\author[T. G. Bisbas et al.]{T. G. Bisbas$^{1}$, T. A. Bell$^{2}$, S. Viti$^{1}$, M. J. Barlow$^{1}$, J. Yates$^{1}$, and M. Vasta$^{1}$\\
$^{1}$ Department of Physics and Astronomy, University College London, Gower Street, London WC1E 6BT, U.K.\\
$^{2}$ Centro de Astrobiolog\'ia (CSIC-INTA), Torrej\'on de Ardoz, 28850 Madrid, Spain}
\date{Accepted , Received ; in original form \today}

\def\cii{[C\,{\sc ii}]\,$158\mu{\rm m}$}
\def\ci{[C\,{\sc i}]\,$609\mu{\rm m}$}
\def\oib{[O\,{\sc i}]\,$146\mu{\rm m}$}
\def\oia{[O\,{\sc i}]\,$63\mu{\rm m}$}
\def\oia{$[$O~{\sc i}$]\,63\mu{\rm m}$}

\begin{document}

\maketitle

\begin{abstract}
We present a model of the photodissociation regions of NGC 4038, which is part of the Antennae galaxies. We have considered one-dimensional slabs of uniform density all having a maximum $A_V=10\,{\rm mag}$, interacting with plane-parallel radiation. The density range in our simulations spans four orders of magnitude ($100\le n\le 10^6\,{\rm cm}^{-3}$) and the UV field strength spans more than three orders of magnitude ($10\le\chi\le10^{4.5}$ multiples of the ${\rm Draine}$ field), from which we generated a grid of about 1400 simulations. We compare our results with {\it Herschel} SPIRE-FTS, CSO and {\it ISO}-LWS observations of eight CO transition lines ($J=1-0$ to $8-7$) and the {\ci} and {\oib} fine structure lines. We find that the molecular and atomic emission lines trace different gas components of NGC 4038, thus single emission models are insufficient to reproduce the observed values. In general, low-$J$ CO transition lines correspond to either low density regions interacting with low UV field strengths, or high density regions interacting with high UV field strengths. Higher $J$ CO transition lines are less dependent on the UV field strength and are fitted by gas with density $n\sim10^{4.5}-10^{5.2}\,{\rm cm}^{-3}$. We find that the observed fine structure line ratio of {\ci}/{\oib} is reproduced by clouds subject to weaker UV fields compared to the CO lines. We make estimates of the $X_{\rm CO}$ factor which relates the CO emission with the column density of molecular hydrogen, and find that it is less than the canonical Milky Way value.
\end{abstract}

\begin{keywords}
astrochemistry -- radiative transfer -- methods: numerical -- galaxies: ISM -- photodissociation region (PDR).
\end{keywords}

\section{Introduction}
There is growing evidence that the properties of the gas in the nuclei of starburst galaxies may be very different from those seen in Galactic star-forming regions and that a high kinetic temperature in the molecular gas may lead to a non-standard initial mass function in the next generation of stars \citep{Spin92}. Observations of gas-phase atoms and molecules in starburst galaxies allow us, in principle, to study the physical characteristics of environments that are quite different from those found in our own Galaxy, e.g. lower or higher abundances of heavy elements, stronger radiation fields and different cosmic-ray ionisation rates. 

Starbursts can occur in disk galaxies, and irregular galaxies often exhibit knots of starburst activity, which can be spread throughout their ISM. Infrared photometry and, later, infrared spectroscopy have provided powerful diagnostics to distinguish between the main emission mechanisms in starburst galaxies. These galaxies are characterised by uncommonly high star formation efficiencies (SFE=$M_{\rm gas}$/SFR), but it remains unclear what physical conditions in the molecular gas produce such high efficiency. Invariably, high star formation efficiency is associated with high column densities of molecular material \citep[e.g. the Kennicutt-Schmidt law;][]{Schm59, Kenn98}. Much of the interest in starburst galaxies arises from the fact that some galaxies, and often very small regions within their nuclei, manage to effectively convert a large amount of gas into stars in a very short time. Often there is plenty of molecular gas present (inferred via its CO emission), so it is not a fuelling question so much as a collection puzzle. It is important to take advantage of the bright emission lines in these types of galaxies in order to investigate the mechanisms that are taking place in the regions they trace.

Emission lines of different coolants, such as {\cii}, {\oia}, {\oib} and CO rotational transitions, are widely used to determine the gas properties. In particular, {\cii} is the dominant cooling line in the warm neutral ISM \citep{Dalg72}. It emits predominantly from the surface layers of photodissociation regions (PDRs), where the far-ultraviolet (FUV) radiation from the HII region is impinging. Oxygen has a slightly higher ionisation potential than hydrogen and therefore is principally in atomic form within neutral regions. In particular, {\oia} is emitted primarily from warm and dense gas and is the main coolant in PDRs after {\cii}. Since {\cii} and both [O\,{\sc i}] lines are emitted from the surfaces of PDRs, their ratios provide a good diagnostic for the strength of the FUV field. However, the {\oia} line suffers from self-absorption, the amount of which is uncertain and depends on the properties of the PDR \citep[i.e. its density structure, the strength of FUV radiation, cosmic-ray ionisation rate, etc.; ][]{Vast10}. 

The Antennae Galaxies, NGC 4038/9 (Arp 244), are one of the best-known examples of interacting starburst galaxies. The two nuclei and the dust obscured overlap region in between exhibit one of the most stunning examples of starburst activity in the nearby universe \citep{Bran09}. With a luminosity just below 10$^{11}$ $L_\odot$ \citep[$\simeq7.5\times10^{10}\,L_{\odot}$,][]{Gao01}, the Antennae do not qualify as luminous infrared galaxies (LIRGs), however they maintain a high star formation rate. The distance to the Antennae has been a subject of recent controversy: although the most commonly assumed value is 19.2 Mpc \citep{Whit99}, more recent observations by \citet{Schw08} have increased the distance estimate to $22\pm3$ Mpc.

\citet{Schu07} studied the properties of the interstellar medium (ISM) in the Antennae galaxies using observations of several low-$J$ CO emission lines. They found that the two nuclei, NGC 4038/9, are both regions of high star formation rate. NGC 4038 is likely to be a streamer caused by the galaxy interaction and it seems to be less disturbed by the collision than NGC 4039. Of particular interest is the `interaction region' (IAR) or `overlap region' which connects NGC 4038/9 and which appears to be optically thick, containing high density regions. Probably most of the star formation activity takes place here, as revealed by {\it Herschel}-PACS observations \citep{Klaa10}, and at an even higher rate than in the nuclei of NGC 4038/9. The bulk of the current star formation in the interacting system, which extends over tens of kpc, is confined to just two compact regions \citep{Bran09}. Usually in a starburst situation, a predominant fraction of the PAH emission arises from the diffuse ISM, heated only by the interstellar radiation field, with only a small fraction arising from PDRs \citep{Drai07}. However, in colliding starburst galaxies such as the Antennae, the situation is reversed, with most of the PAH emission coming from the regions of recent star formation, as in the map shown by \citet{Bran09}, where PAH emission primarily traces the PDRs and hence the environment of OB clusters. The nuclei of both NGC 4038 and NGC 4039 are thus very active star-forming regions, surrounded by regions of moderately active star formation. Finally, while the nucleus of NGC 4039 appears brighter in the [Ne\,{\sc iii}] and [S\,{\sc iv}] emission lines, the lack of [Ne\,{\sc v}] emission and the low [O\,{\sc iv}]/[Ne\,{\sc ii}] ratio rules out the presence of an AGN for both nuclei \citep{Bran09}.

Maps of the Antennae galaxies in the {\cii} cooling line, at an angular resolution of 55\arcsec, showed that the starburst activity is confined to small regions of high SFE \citep{Niko98}. \citet{Wils03} found an excellent correlation between the strengths of the CO emission and the 15$\mu$m broad band emission seen by {\it ISO}, and determined masses of 3--6$\times$10$^{8}$~$M_{\odot}$ for the largest molecular complexes, typically an order of magnitude larger than the largest structures found in the disks of more quiescent spiral galaxies. 

Recently, Schirm et al. (2013, \emph{in press}; hereafter `S13') presented {\it Herschel} SPIRE-FTS observations of five CO transition lines ($J=4$--3 to $J=8$--7), both [C\,{\sc i}] transitions and the [N\,{\sc ii}] 205$\mu$m transition. Using radiative transfer analysis they found that a one-component model cannot represent a physical solution, since it only recovers warm kinetic temperature ($T_{\rm kin}>100\,{\rm K}$) and low density ($n_{{\rm H}_2}<10^3\,{\rm cm}^{-3}$) molecular gas. Instead, they find evidence for a two-component model with an additional cold ($T_\mathrm{kin}\sim10$--30 K) molecular gas component. Using PDR models performed by \citet{Holl12}, they found that in both NGC 4038/9 nuclei, as well as in the overlap region, their emission line observations were best fit by models with an average interstellar radiation field strength of about $\chi\sim10^{2}$--$10^{3}\chi_{0}$, where $\chi_{0}$ is the standard \citet{Drai78} field. 

The work by \citet{Karl10} using high resolution N-body/Smoothed Particle Hydrodynamics simulations produced remarkably similar morphological density structures to those observed. Using dust radiative transfer calculations, they also found a good agreement with the observed $70\,\mu{\rm m}$, $100\,\mu{\rm m}$ and $160\,\mu{\rm m}$ emission \citep{Karl13}. \citet{Schu07} compared observations of the $^{12}$CO(1--0), (2--1), (3--2) and {\cii} lines with PDR models. They found good agreement when the modelled clouds had moderately high densities, up to $4\times10^4$~cm$^{-3}$, and rather low kinetic temperature ($\le25$~K). This allowed them to estimate the total molecular gas mass of the Antennae to be $\sim10^{10}$ $M_\odot$.

However, several authors have considered the presence of various heating mechanisms other than UV radiation. \citet{Fisc96} argued that in the nucleus of NGC 4039 the observed H$_2$ 1--0 S(1) line is primarily emitted by C-shocks and not by the UV-exposed gas in PDRs. \citet{Wils00} presented two scenarios based on CO observations to explain the strong mid-IR continuum emission seen in some `super-giant molecular complexes', namely: i) the presence of young star-forming sites in which the embedded O stars have not yet managed to blow away the surrounding dust; and ii) the presence of shocked boundary layers of colliding clouds. Furthermore, \citet{Schu07} found that apart from shock-driven heating, cosmic rays and ambipolar diffusion could potentially increase the emission of the observed lines, while S13 argue that the mechanical heating by supernovae and turbulence is probably sufficient to match the total cooling and heat up the molecular gas in the Antennae. Recently, \citet{Herr12} using ALMA CO(3--2) interferometry and VLT/SINFONI imaging spectroscopy of H$_2$ 1--0 S(1) found that the emission of molecular hydrogen is powered by shocks and that those two lines can be used respectively as tracers for energy dissipation and gas mass.

In this paper we present results from a suite of PDR models computed using the {\sc 3d-pdr} code \citep{Bisb12} and we compare our models with ground-based, {\it ISO}-LWS and the recent {\it Herschel} SPIRE-FTS observations of the NGC 4038 nucleus. We investigate the correlation between the local density and the UV field strength in different CO transition lines and fine structure lines. While previous studies have suggested that PDRs cannot be the only source of emission lines and that additional heating mechanisms are needed, we explore whether this claim still holds if one relaxes the assumption of a single-component model. In Section \ref{sec:numeric} we discuss the numerical treatment we follow for the PDR models. In Section \ref{sec:observ} we describe the observations of NGC 4038 taken from the literature and our methodology for convolving them to a common beam size. In Section \ref{sec:results} we present the results from our grid of PDR models and in Section \ref{sec:disc} we provide further discussion on our outcomes. We conclude in Section \ref{sec:conc}.

\section{Numerical treatment}\label{sec:numeric}
We use {\sc 3d-pdr} \citep{Bisb12}, a three-dimensional time-dependent astrochemistry code designed for treating PDRs of arbitrary density distribution. {\sc 3d-pdr} has been fully benchmarked with other PDR codes following the tests presented in \citet{Roll07} and has been previously used to study the chemistry in turbulent star-forming clouds \citep{Offn13}. For full technical details about the code, see \citet{Bisb12}.

We consider one-dimensional uniform density clouds following the same methodology as described in \citet{Bisb12}. In particular, we place 20 elements (depth points) in $A_V$ dex intervals, starting at ${\rm A}_{V,{\rm min}}=10^{-5}\,{\rm mag}$. The total visual extinction in all clouds is fixed at ${\rm A}_{V,{\rm max}}=10\,{\rm mag}$. Thus, each cloud consists of ${\cal N}_{\rm elem}=120$ elements distributed logarithmically with increasing cloud depth. The elements of the cloud are aligned with two opposing {\sc healpix} rays \citep{Gors05} while we assume very high optical depths for all other ray directions, implying that each cloud is a semi-infinite one-dimensional slab. In this setup, convergence of the various model properties is treated as described in \citet{Bisb12}. In all cloud models discussed below, we note that the temperatures, abundances and line emissions are calculated up to the maximum $A_V=10\,{\rm mag}$, however we examine the resulting integrated intensities (surface brightnesses) up to smaller values in visual extinction.

The chemical network we use in this paper is a subset of the most recent UMIST data base of reaction rates \citep[UMIST 2012; ][]{Mcel13}. This ``reduced'' chemical network consists of 33 species (including electrons) and 320 reactions and is used to model the gas-phase chemistry. It includes the four most abundant elements H, He, O, and C, as well as Mg. We include Mg to represent the metal contribution in the ionization fraction. It neglects polycyclic aromatic hydrocarbons (PAHs), although they have been considered for the photoelectring heating. We use the UMIST 2012 rates with the exception of the reaction rates for the photodissociation of H$_2$ and CO and the photoionisation of carbon. For these particular cases we adopt the treatment of \citet{Lee96} and \citet{vanD88}, and \citet{Kamp00}, respectively. 

Following a recent update of the {\sc 3d-pdr} code, we calculate the rate of molecular hydrogen formation on grains using the detailed treatment of \citet{Caza02, Caza04, Caza10}. The dust temperature for each cloud element is determined using the methodology of \citet{Holl91} based on heating due to FUV photons incident on the cloud surface. We have assumed that dust is made of a mix of silicate and carbonaceous grains. The grain radius, $r_{\rm d}$, is taken to be $10^{-7}\,{\rm cm}$, the grain number density, $n_{\rm d}$, is $2\cdot10^{-12}\times n\,{\rm cm}^{-3}$ where $n$ is the gas density, and the gas-to-dust mass ratio, $M_{\rm g}/M_{\rm d}$, is $100$. The turbulent velocity resulting from energy dissipation through shocks, which governs the turbulent heating and escape of cooling line emission within the cloud, is set to $v_{\rm turb}=1.5$~km\,s$^{-1}$, typical of individual molecular clouds \cite[e.g.,][]{Holl99}. Further details of the model parameters can be found in \citet{Bisb12}.
As shown by \citet{Baye11}, the cosmic-ray ionization rate per H$_2$ molecule, $\zeta_{\rm CR}$, plays a crucial role in determining the chemistry and the thermal balance in PDRs. It is a dominant factor controlling the cosmic-ray-driven chemistry and heating and has a significant effect in the innermost part of the cloud. In this paper however, we do not explore $\zeta_{\rm CR}$ as a free parameter. Instead we assume it to be constant with a value of $\zeta_{\rm CR}=5\times10^{-17}\,{\rm s}^{-1}$. A higher value of this rate would increase the temperature of the inner part of PDRs. {\sc 3d-pdr} performs an iterative scheme to determine the gas temperature by balancing the heating and cooling rates in each cloud element. We evolve the chemistry in each simulation for 10 Myr, by which time it has reached equilibrium \citep{Baye09}. This timescale is smaller than the 100Myr commonly used, however due to the initial conditions adopted here we reach equilibrium in a much shorter timescale.

\citet{Vazq05} used Starburst99 \citep{Leit99} models to fit the cluster colour-colour diagrams for NGC 4038/9 and found that the system is best reproduced assuming solar chemical composition. Conversely, they found that models with subsolar chemical composition can clearly be excluded. Direct abundance estimates based on nebular electron temperature diagnostics have not been published for the Antennae galaxies. X-ray spectra have indicated abundances in many regions that are consistent with solar \citep{Bald06} so we adopt the solar abundances of \citet{Aspl09}, a summary of which is shown in Table \ref{tab:abund}. We note that these abundances do not correspond to the standard Milky Way gas-phase abundances but rather to undepleted solar abundances.

\begin{table}
\caption{Initial gas-phase elemental abundances relative to total hydrogen nuclei used in the {\sc 3d-pdr} code for the grid of models \citep{Aspl09}}
\centering
\label{tab:abund}
\begin{tabular}{l l l l}
\hline
H     & $4\times10^{-1}$   & Mg$^+$ & $3.98\times10^{-5}$ \\
H$_2$ & $3\times10^{-1}$   & C$^+$  & $2.69\times10^{-4}$  \\
He    & $8.5\times10^{-2}$ & O      & $4.90\times10^{-4}$ \\
\hline
\end{tabular}
\end{table}

The density range we examine spans four orders of magnitude, where $10^2\le n\le10^6\,{\rm cm}^{-3}$. For densities less than $10^2\,{\rm cm}^{-3}$ the medium can be considered ionized and can only be simulated in detail using photoionization codes such as {\sc mocassin} \citep{Erco03,Erco05} and {\sc torus} \citep{Hawo12}. For densities greater than $10^6\,{\rm cm}^{-3}$, the medium can be considered fully molecular, as the UV radiation is attenuated very rapidly. The range of incident UV radiation field intensities we consider spans more than three orders of magnitude, $10\le{\chi}/{\chi_{0}}\le10^{4.5}$, where $\chi_{0}$ is the standard \citet{Drai78} interstellar radiation field. For each of the two ranges of density and UV radiation field we run 10 simulations per dex. We thus perform a total grid of 1400 simulations using the initial conditions described above. 

\section{Observational data}\label{sec:observ}

We use observational data of various lines taken from different instruments. The CO(1--0) line is a SEST observation \citep{Aalt95} and the lines of CO(2--1), (3--2), (4--3) and (7--6), as well as {\ci}, are CSO observations \citep{Baye06}. The CO(5--4) and (8--7) lines and additional CO(4--3), (6--5) and (7--6) lines are {\it Herschel} SPIRE-FTS observations (see S13). Finally, we use {\oia}, {\oib} and {\cii} {\it ISO}-LWS observations \citep{Fisc96}. 

Since the observations were taken using different beam sizes, we convolve all line intensities to the same angular resolution. We use two different resolutions: the 43\arcsec of the {\it Herschel} SPIRE-FTS beam when comparing CO lines only, and the 80\arcsec beam of {\it ISO}-LWS when comparing the CO and fine structure lines. For the convolution to the final line intensities, we use the scaling factor for elliptical sources, which depends on the source size and the initial and final spatial resolution, given by
\begin{eqnarray}
\label{eqn:conv}
{\rm fact}_1=\sqrt{\frac{\theta_{\rm mb,init}^2+FWHM_{s,x}^2}{\theta_{\rm mb,fin}^2+FWHM_{s,x}^2}}\sqrt{\frac{\theta_{\rm mb,init}^2+FWHM_{s,y}^2}{\theta_{\rm mb,fin}^2+FWHM_{s,y}^2}}
\end{eqnarray}
where FWHM$_{s,x}=13.2\arcsec$ and FWHM$_{s,y}=9.9\arcsec$ are the assumed dimensions of the source region. The value of $\theta_{\rm mb,init}$ is the initial beam size and $\theta_{\rm mb,fin}$ is the final (convolved) beam size. We then convert the resultant convolved line intensities to units of ${\rm erg}\,{\rm cm}^{-2}\,{\rm s}^{-1}\,{\rm sr}^{-1}$ by multiplying them by
\begin{eqnarray}
{\rm fact}_2=1.02\times10^{15}\cdot\nu_{ij}^3
\end{eqnarray}
where $\nu_{ij}$ is the transition frequency in GHz. For the {\oia}, {\oib} and {\cii} lines observed by {\it ISO}-LWS we use the standard factor 
\begin{eqnarray}
{\rm fact}_3=\frac{\pi (80\arcsec/2)^2}{206264.8^2}\,
\end{eqnarray}
to convert the fluxes to units of ${\rm sr}^{-1}$. A summary of all the observational data we use are presented in Table \ref{tab:observ}. 


\begin{table*}
\caption{Summary of SEST, {\it ISO}-LWS, CSO, and {\it Herschel} SPIRE-FTS line observations. All observations are for NGC 4038. The critical densities are all for $T=100\,{\rm K}$ and those for the fine structure lines correspond to the collisions with ortho-H$_2$.}
\label{tab:observ}
\begin{tabular}{c l c c c cc c l}
\hline
\hline
ID&Line&Frequency&$n_{\rm crit}$&$\theta_{\rm mb,init}$&\multicolumn{2}{c}{Intensity}&Reference&Comments\\
& &(GHz)& $(10^3\times{\rm cm}^{-3})$ & &\multicolumn{2}{c}{($\times 10^{-6}\,{\rm erg}\,{\rm cm}^{-2}\,{\rm s}^{-1}\,{\rm sr}^{-1}$)}& & \\
& &     & & &$\theta_{\rm mb,fin}=80\arcsec$&$\theta_{\rm mb,fin}=43\arcsec$& & \\
\hline
 1  & CO (1-0)	& 115.27  &  1.8   & 43\arcsec    & $0.01\pm0.001$& $0.03\pm0.003$ & $a$   & SEST	\\
 2  & CO (2-1)  & 230.54  &  9.7   & 30.5\arcsec  & $0.07\pm0.02$ & $0.24\pm0.06$  & $b$   & CSO	\\
 3  & CO (3-2)  & 345.80  &  32.4  & 21.9\arcsec  & $0.15\pm0.01$ & $0.49\pm0.02$  & $b$   & CSO	\\
 4a & CO (4-3)  & 461.04  &  80.6  & 14.55\arcsec & $0.24\pm0.02$ & $0.79\pm0.06$  & $b$   & CSO	\\
 4b & 		& 	  &        & 43\arcsec    & $0.49\pm0.03$ & $1.60\pm0.10$  & $c$   & {\it Herschel} 	\\
 5  & CO (5-4)  & 576.28  &  154.6 & 43\arcsec    & $0.27\pm0.02$ & $0.88\pm0.08$  & $c$   & {\it Herschel} 	\\
 6  & CO (6-5)  & 691.47  &  263.8 & 43\arcsec    & $0.26\pm0.03$ & $0.84\pm0.10$  & $c$   & {\it Herschel} 	\\
 7a & CO (7-6)  & 806.65  &  407.4 & 8.95\arcsec  & $0.09\pm0.02$ & $0.28\pm0.07$  & $b$   & CSO  		\\
 7b &   	&         &        & 43\arcsec    & $0.13\pm0.03$ & $0.43\pm0.11$  & $c$   & {\it Herschel}      \\
 8  & CO (8-7)  & 921.80  &  604.0 & 43\arcsec    & $0.17\pm0.02$ & $0.56\pm0.08$  & $c$   & {\it Herschel}      \\
 9  & {\ci}     & 492.16  &  1.1 & 14.55\arcsec & $0.09\pm0.01$ & --             & $b,d$ & CSO	        \\
10  & {\oia}    & 4758.61 &  $6.4\times10^3$ & 80\arcsec    & $44.0\pm0.93$ & --             & $e$   & {\it ISO}-LWS       \\
11  & {\oib}    & 2067.53 &  $5.8\times10^3$ & 80\arcsec    & $1.79\pm0.35$ & --             & $e$   & {\it ISO}-LWS       \\
12  & {\cii}    & 1897.42 &  4.5 & 80\arcsec    & $31.3\pm0.43$ & --             & $e$   & {\it ISO}-LWS	\\

\hline
\end{tabular}
\begin{center}
$^a$ \citet{Aalt95}, $^b$ \citet{Baye06}, $^c$ Schirm et al. (2013; \emph{in press}), $^d$ \citet{Geri00}, $^e$ \citet{Fisc96}
\end{center} 
\end{table*}

\section{Results}\label{sec:results}

\begin{figure}
\begin{center}
\includegraphics[width=9cm]{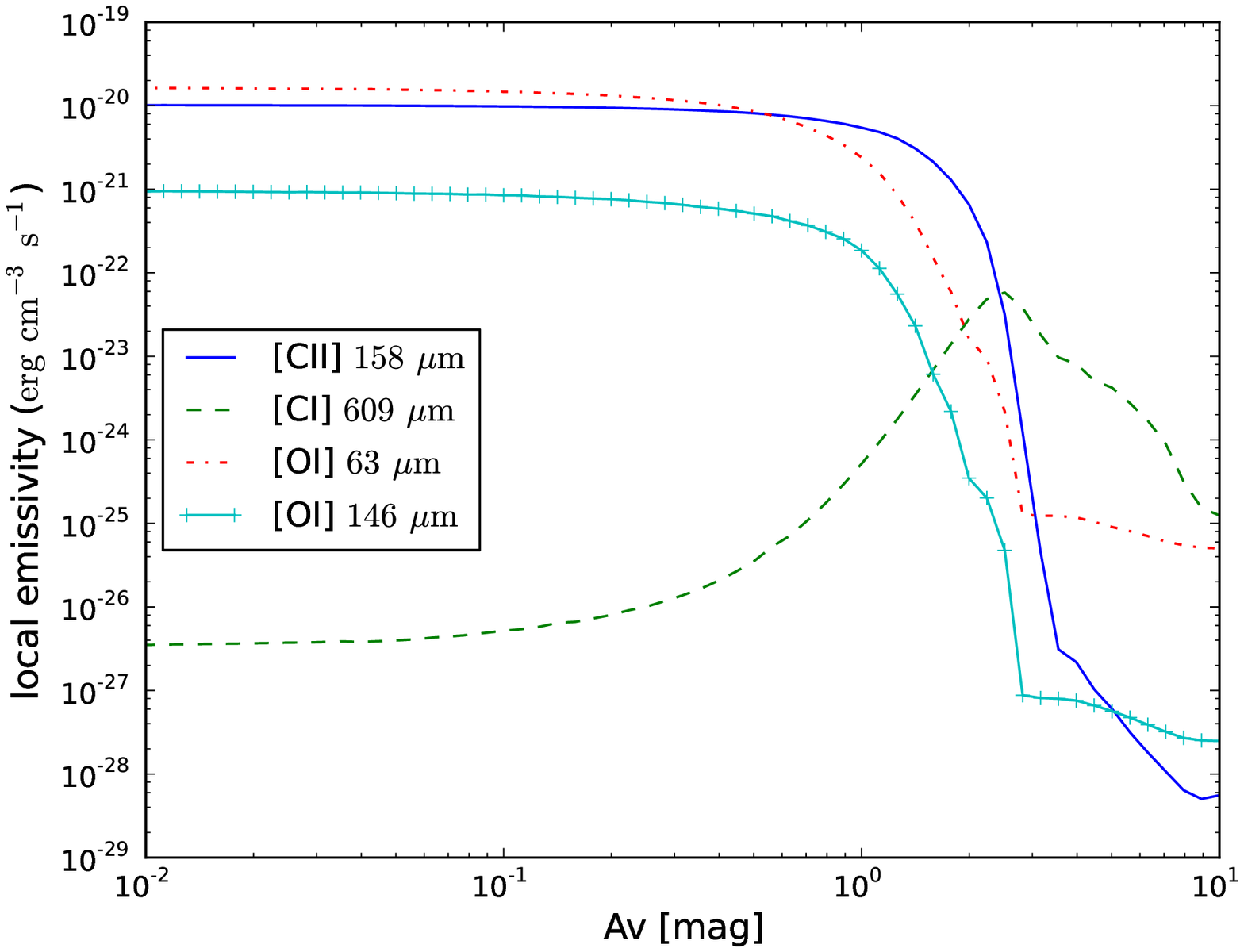} 
\includegraphics[width=9cm]{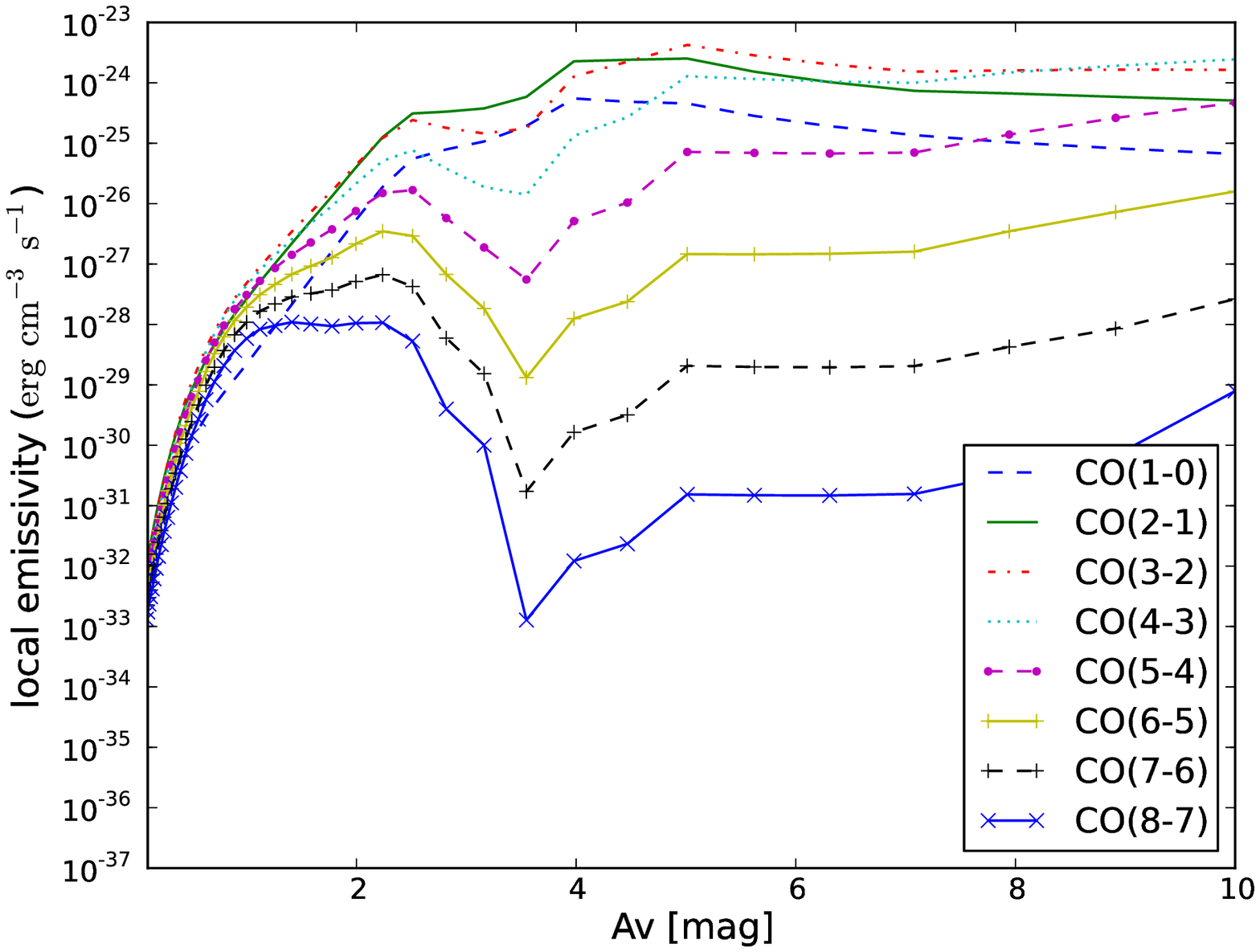} 
\caption{Top panel: local emissivities of {\cii}, {\oia}, {\oib}, and {\ci} for a PDR of uniform density of $n=10^3\,{\rm cm}^{-3}$ interacting with a plane-parallel UV field of strength $\chi=10^{3.5}\chi_{0}$. {\cii}, {\oia} and {\oib} are significantly weaker at ${\rm A}_V\gtrsim2\,{\rm mag}$ indicating that they are good tracers for the UV radiation close to the HII region, while {\ci} has a maximum peak at ${\rm A}_V\sim3\,{\rm mag}$ i.e. at intermediate depths. Bottom panel: local emissivities of CO with $J=(1-0)..(8-7)$ versus ${\rm A}_V$. For low-$J$ transitions i.e. $J\ge4$ we see that the maximum peak comes from the innermost part of the PDR i.e. ${\rm A}_V\gtrsim5\,{\rm mag}$ indicating that they are good diagnostics for examining the conditions close to molecular regions. For higher $J$ there is a local peak at ${\rm A}_V\lesssim4\,{\rm mag}$ which indicates that there is some significant contribution from the PDR parts closer to the ionized region.}
\label{fig:loc}
\end{center}
\end{figure}

In order to first determine how the transition lines of the different coolants can be used as diagnostics to study the PDR properties, we consider a simple model of a uniform density cloud with a total H-nucleus number density of $n=10^3\,{\rm cm}^{-3}$ interacting with a UV field of strength $\chi=10^{3.5}\chi_{0}$. We plot the various line emissivities as a function of cloud depth in Fig.~\ref{fig:loc}. The top panel shows the local emissivity versus visual extinction $A_V$ for the {\cii}, {\oia}, {\oib}, and {\ci} fine structure lines. We see that the maximum local emissivity for the first three lines comes from the part of the PDR closest to the ionised medium. In addition, at $A_V \gtrsim 2$ mag they abruptly decrease, implying that their emission is governed by the UV radiation field, thus they are good diagnostics for its strength. The local emissivity of the {\ci} line, on the other hand, has a peak at $A_V \sim 3$ mag or less, depending on the physical parameters of the PDR, and is in general emitted from a narrow depth range at intermediate visual extinctions.

The bottom panel of Fig. \ref{fig:loc} shows the emissivities of various CO rotational lines. Here we see that the low-$J$ CO lines have their emission peak at $A_V \gtrsim 5$ mag, therefore the source of their emission is primarily deeper within the molecular cloud. On the other hand, high-$J$ CO lines appear to have two peaks, the first of which is at $A_V \lesssim 2$ mag, thus their integrated emission has a significant contribution from the PDR closer to the ionised medium. It is interesting to note that the local emissivities of the high-$J$ transitions also start to rise again with increasing $A_V$ beyond 4 mag. This is because the low-lying transitions begin to have very large optical depths, causing the higher energy levels to become populated deeper into the cloud. This is in part a result of the semi-infinite slab geometry we adopt in our models, which prevents emission from escaping from the back face of the cloud.

In \S\ref{ssec:nclouds} we focus on estimating the best-fit range of densities as well as the number of model clouds needed to reproduce the observed values of the different CO line transitions. In \S\ref{ssec:UVestimate} we focus on estimating the best-fit range of the UV field strength by examining the observed fine structure lines in comparison with our modeled data cubes. In \S\ref{ssec:unified} we combine the results for all the CO transition lines into one unified picture as a function of the density and the UV field strengths. Finally, in \S\ref{ssec:comparison} we compare our findings with those of other authors and in \S\ref{ssec:xfactor} we estimate the $X_{\rm CO}$ factor of NGC 4038 based on the results we have obtained.

\subsection{Constraining the best-fit density range and the number of model clouds}
\label{ssec:nclouds}

Molecular line ratios are often used to constrain the excitation conditions and chemistry of the molecular gas. We therefore use the CO ratios from our grid of models; however since ratios can often be fit by several models, seldom constraining individual parameters (i.e density and especially radiation fields) and moreover since the matched ratios can also be obtained from different absolute intensities, we couple this standard procedure with an analysis of the best-fit intensities for each individual line. We use the 43\arcsec convolved CO lines of Table \ref{tab:observ}. 

\begin{figure*}
\begin{center}
\includegraphics[width=0.48\textwidth]{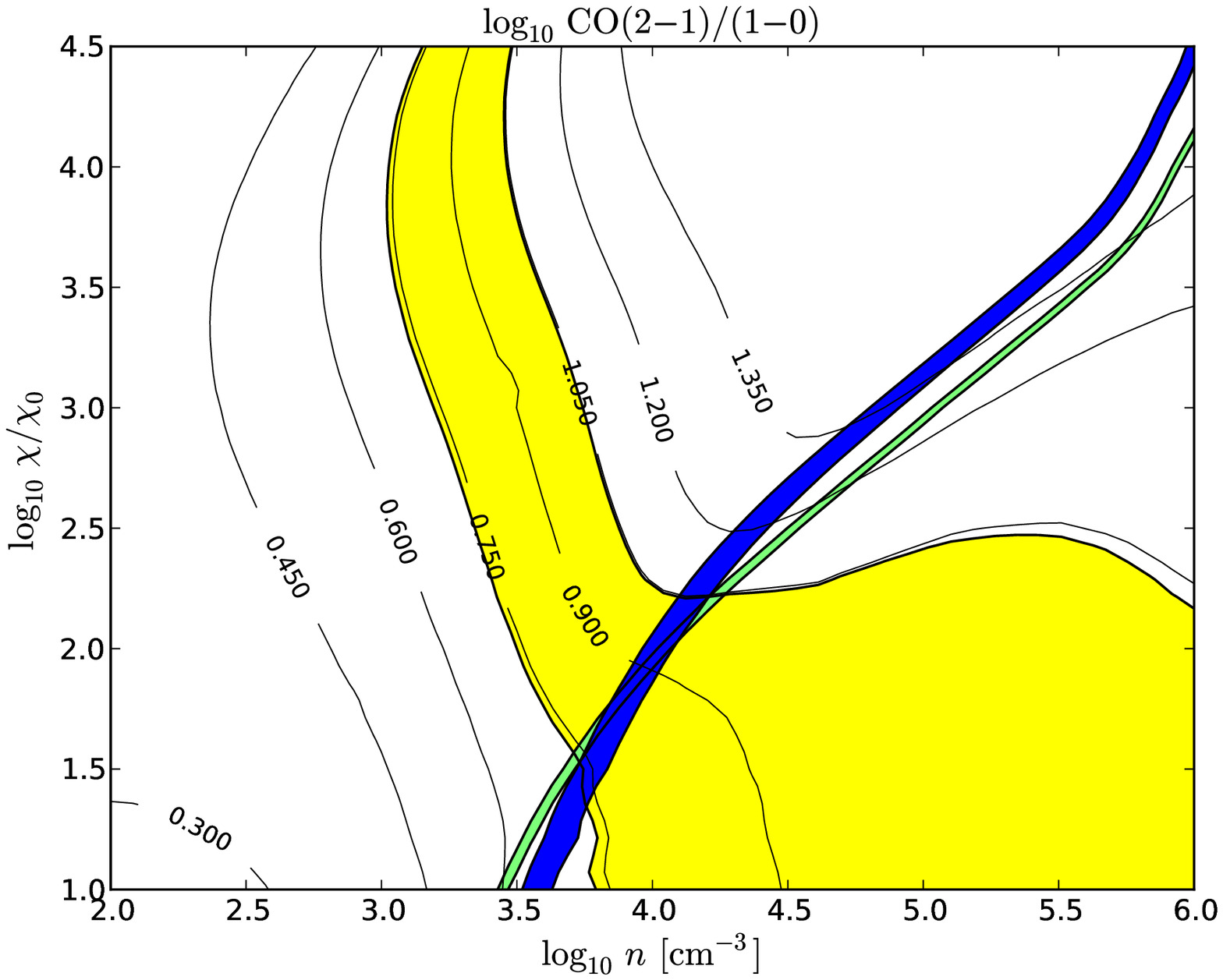} 
\includegraphics[width=0.48\textwidth]{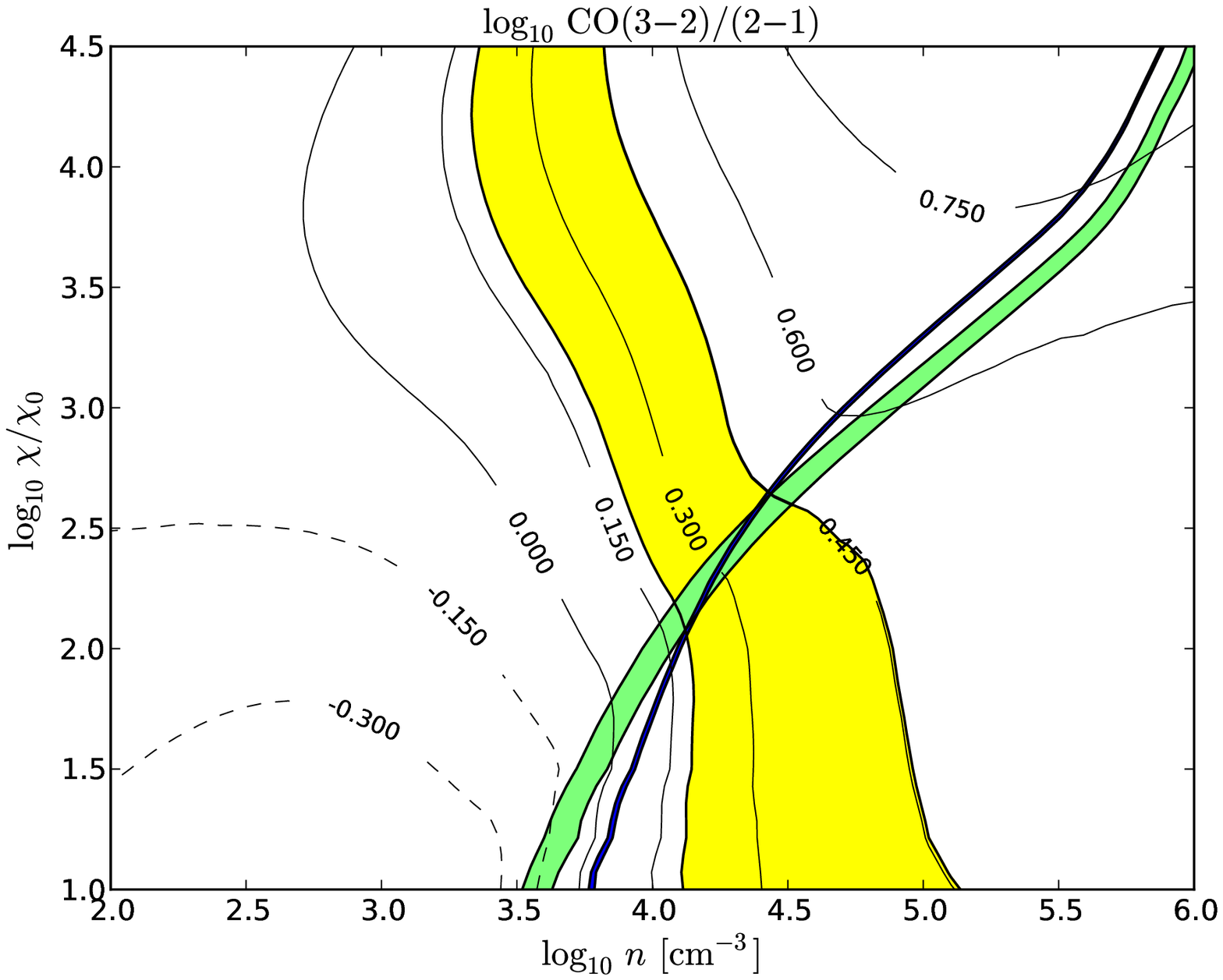} 
\includegraphics[width=0.48\textwidth]{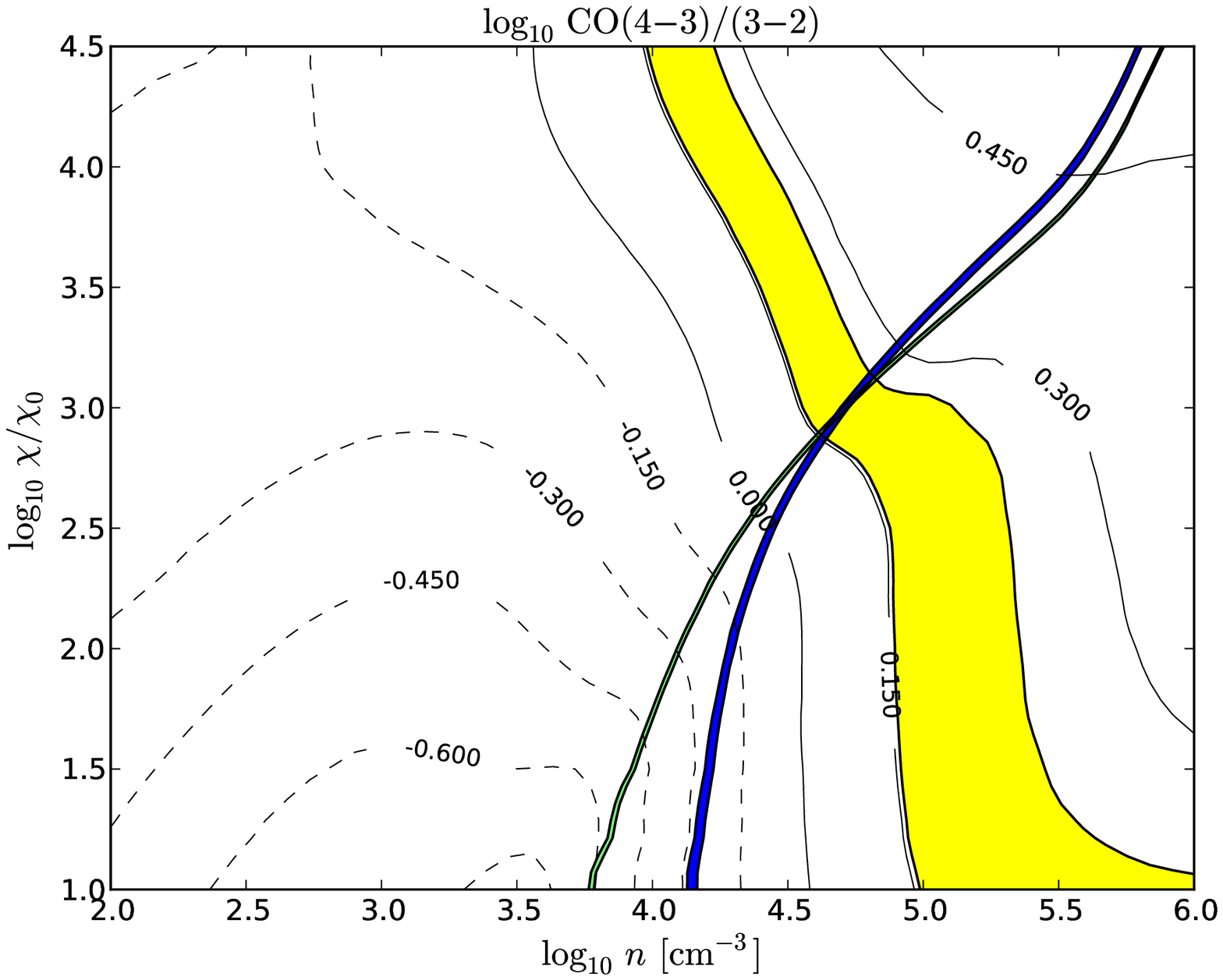} 
\includegraphics[width=0.48\textwidth]{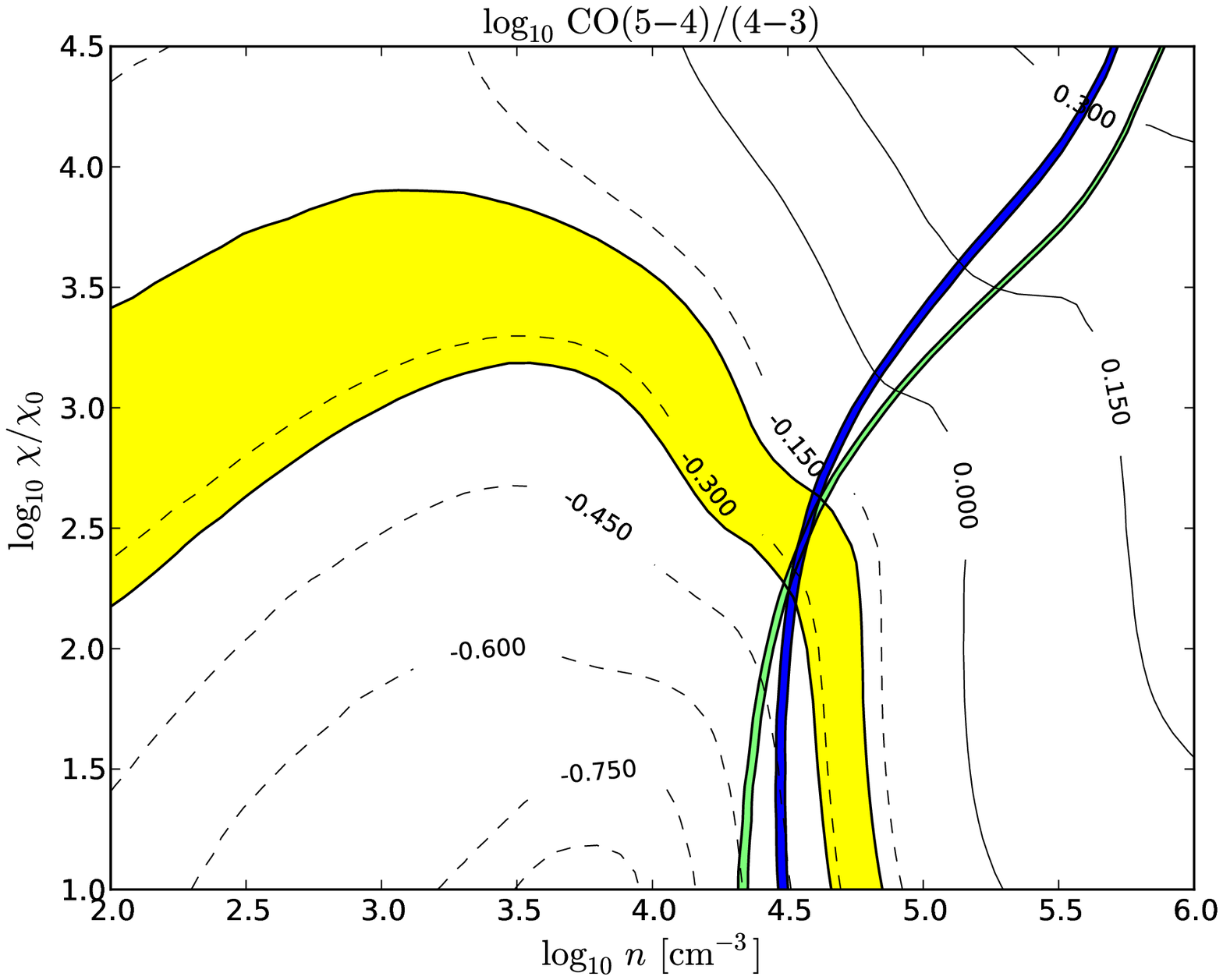} 
\caption{Line ratios and individual line intensities (surface brightnesses) of CO for consecutive $J$ transition lines, where the IDs are those listed in the first column of Table \ref{tab:observ} corresponding to $\theta_{\rm mb,fin}=43\arcsec$. In all panels, yellow colour represents the ratio of the two corresponding lines, green colour the lower $J$ transition of the observed ratio and blue colour the higher $J$ transition of the observed ratio. The black contours correspond to the logarithmic values of the ratio produced by the models. The thickness of the shaded regions corresponds to the observational error given in Table \ref{tab:observ}. Top left panel plots the CO(2-1)/(1-0) ratio i.e. ID=2/1 where the IDs are those of the same Table. Top right panel plots the CO(3-2)/(2-1) ratio i.e. ID=3/2. Bottom left panel plots the CO(4-3)/(3-2) ratio i.e. ID=4a/3. Bottom right panel plots the CO(5-4)/(4-3) ratio i.e. ID=5/4b.}
\label{fig:ratios1}
\end{center}
\end{figure*}

\begin{figure*}
\begin{center}
\includegraphics[width=0.48\textwidth]{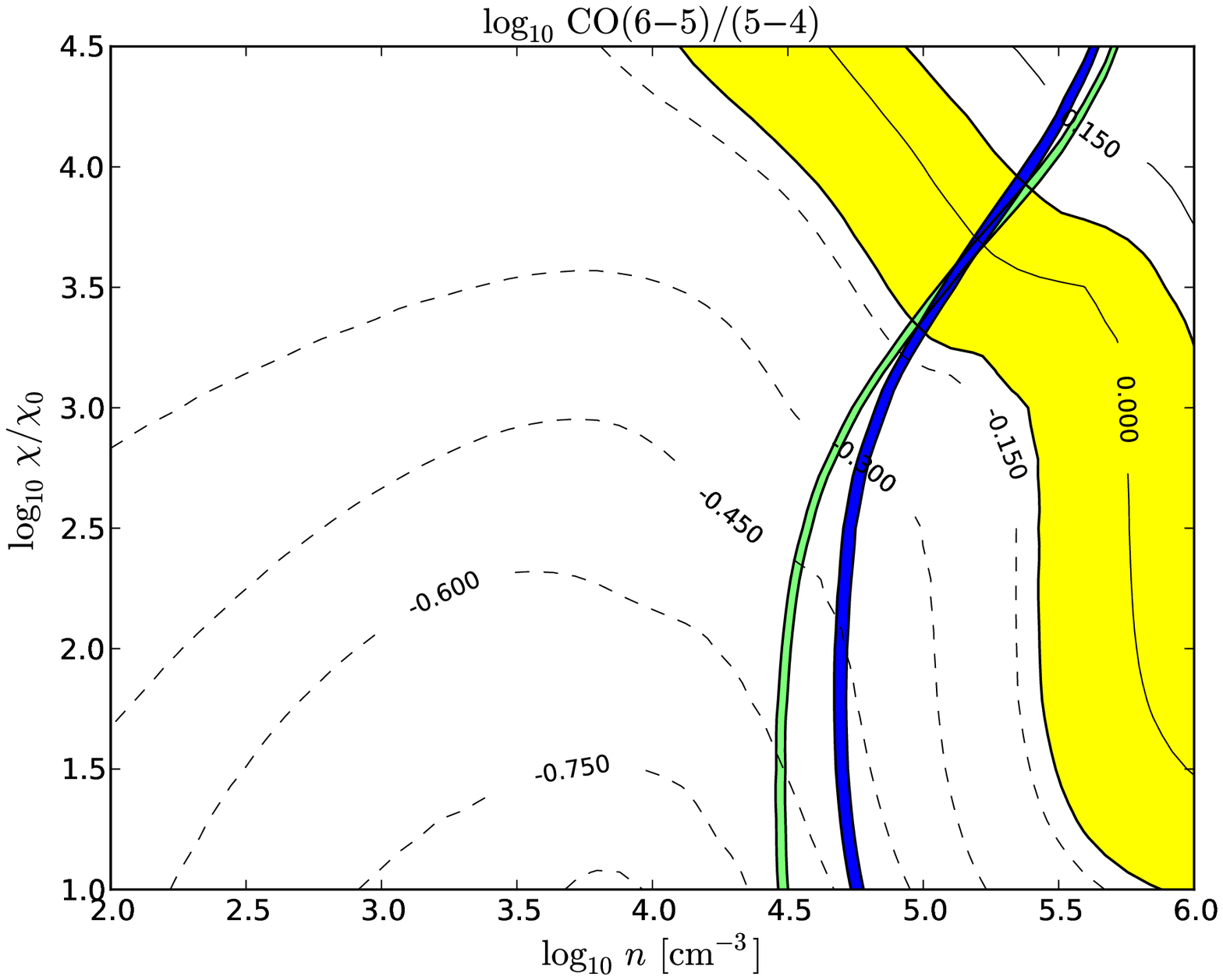} 
\includegraphics[width=0.48\textwidth]{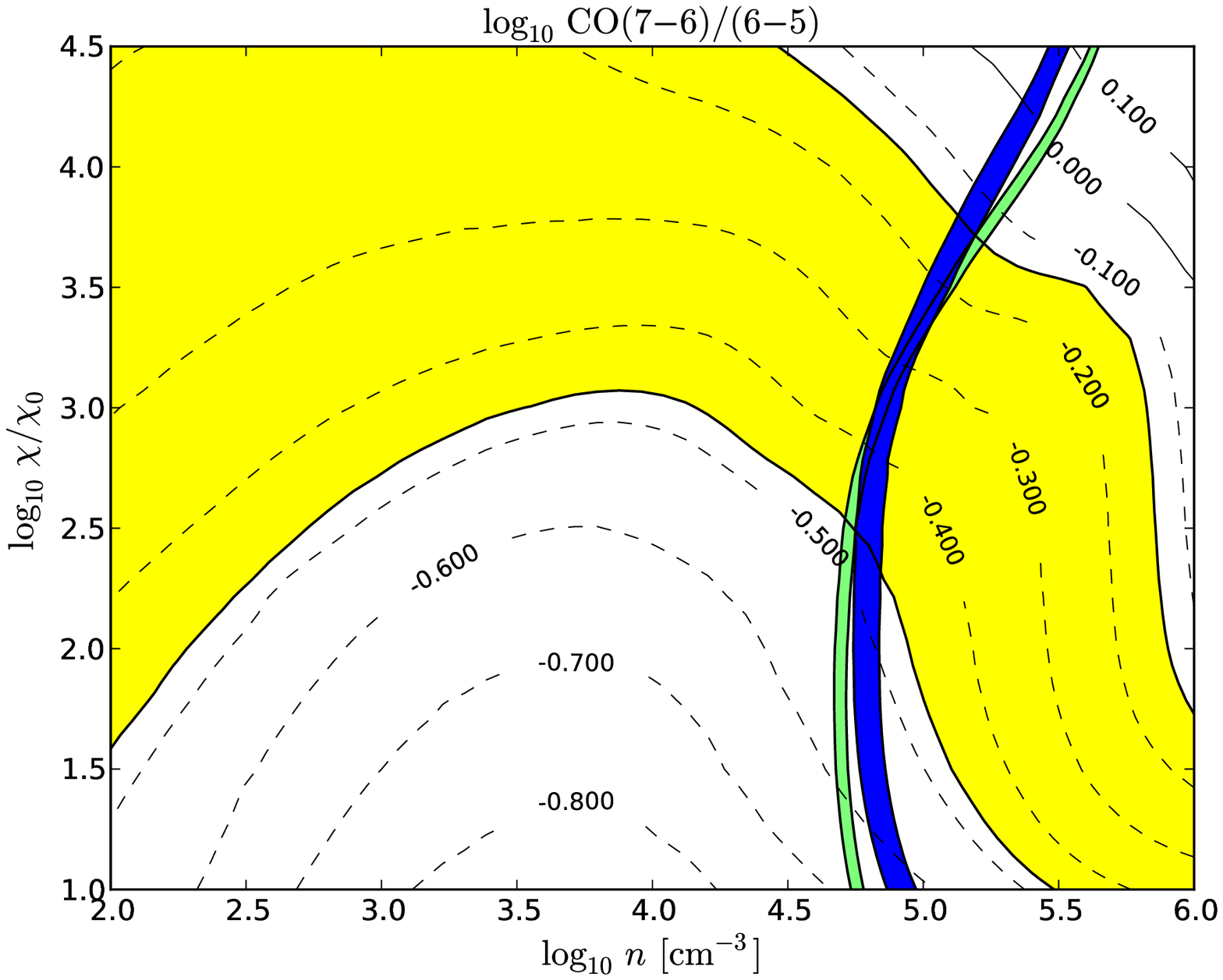} 
\includegraphics[width=0.48\textwidth]{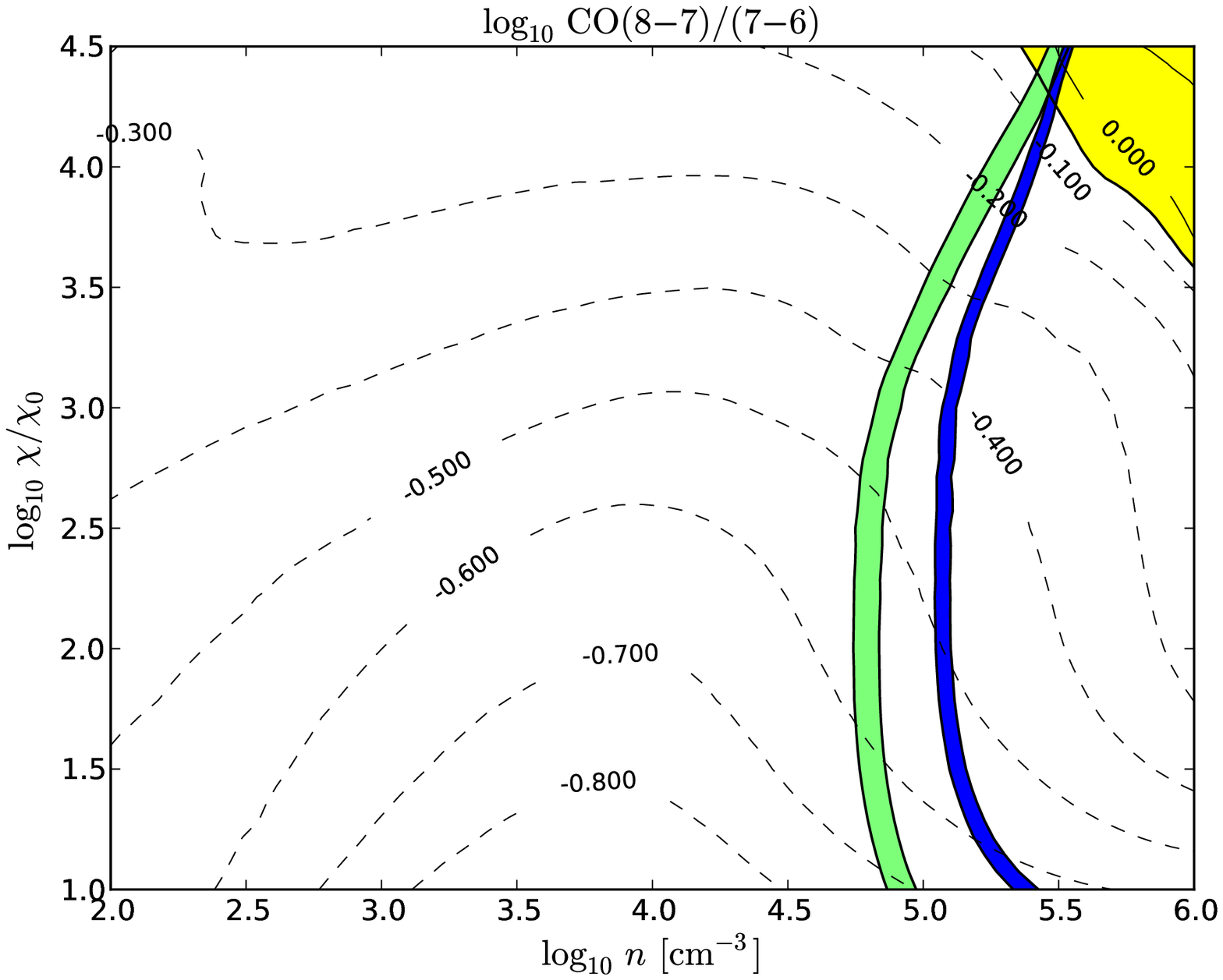} 
\includegraphics[width=0.48\textwidth]{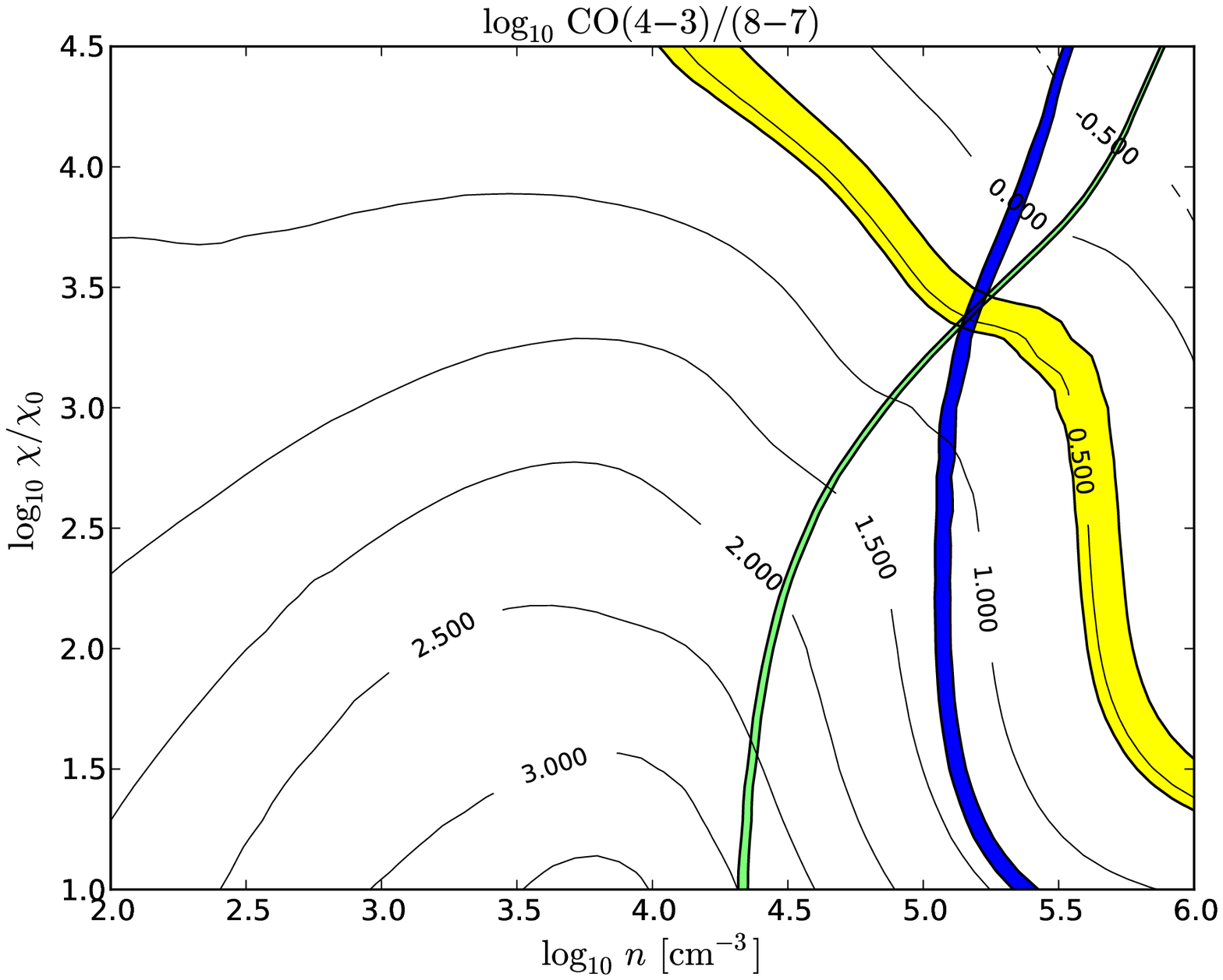} 
\caption{As in Fig. \ref{fig:ratios1}. Top left panel plots the CO(6-5)/(5-4) ratio i.e. ID=6/5. Top right panel plots the CO(7-6)/(6-5) ratio i.e. ID=7a/6. Bottom left panel plots the CO(8-7)/(7-6) ratio i.e. ID=8/7b. Bottom right panel plots the ratio CO(4-3)/(8-7) i.e. ID=4b/8 which corresponds to the lower and higher transition lines of {\it Herschel} SPIRE-FTS instrument respectively. In this panel we additionaly find that there are two `pairs' of best-fit regions corresponding to i) intermediate-densities and high UV field strengths and ii) high densities and low UV field strengths.}
\label{fig:ratios2}
\end{center}
\end{figure*}

Since we assume a source size of $13.2\arcsec\times9.9\arcsec$ when convolving the observed line intensities to a common beam size of $43\arcsec$, we adopt the same implicit beam filling factor as an upper limit when comparing model line intensities to the observed values. For a $43\arcsec$ beam, this upper limit to the filling factor is 0.07, but the best-fit value needed to reproduce the observed intensities may in fact be lower. We then introduce a scaling factor applied to the model cloud integrated intensity in order to match it to the observed line intensity. We refer to this scaling factor as $N_\mathrm{cloud}$, the number of model clouds needed to reproduce the total observed intensity in a particular transition; this factor may be greater or less than 1. In the case that it is greater than 1, it implies that more than one model cloud must be present within the area of the assumed source size in order to reproduce the observed line intensity. Alternatively, if $N_\mathrm{cloud}$ is less than one, this implies that the effective beam filling factor is less than our upper limit of 0.07 and that a single model cloud is capable of reproducing all of the observed emission. In reality, of course, there is degeneracy between the number of clouds that occupy the beam and the beam dilution describing the angular extent of those clouds. Therefore, $N_\mathrm{cloud}$ serves as a combined quantity that captures both properties of the underlying cloud distribution.

The final beam size of the observations is 43\arcsec which given that the distance to the Antenna galaxies is about 20 Mpc, corresponds to roughly 4 kpc of gas; if we were to plot the molecular ratios for an $A_V=10\,{\rm mag}$ slab we would find that, although they could be fitted by several models, the individual line intensities would be larger than the observed ones. This means that less than one molecular cloud would be needed to account for the observations. Assuming a uniform filling factor, it is thus unlikely that the individual clouds are an average of 10 mags in visual extinction. We therefore arbitrarily plot the ratios of the line intensities integrated up to $A_V=2\,{\rm mag}$ only: as one can see from Figs. \ref{fig:ratios1} and \ref{fig:ratios2}, a large range of densities match the observed ratios. In order to calculate $N_{\rm cloud}$, we adopt the following procedure. We find the lowest density which corresponds to the best-fit ratio of two $J-$ transition lines, and the highest density which corresponds to the area where the two individual lines intersect; we consider the absolute line intensities of the individual lines for the above densities; we then estimate the number of clouds one would need for both limits to account for the observed intensities by using the simple relation
\begin{eqnarray}
N_{\rm clouds} = \frac{I_{\rm sim}}{I_{\rm obs}}\,,
\end{eqnarray}
where $I_{\rm obs}$ and $I_{\rm sim}$ correspond to the observed and the calculated intensity respectively. As we have stated above, if less than one cloud is needed i.e the theoretical line intensity is less than the observed one, then the best fit model is incorrect. Of course, a possible explanation for a larger-than-observed intensity is the beam dilution effect i.e. it is possible that the observed intensities are underestimated due to beam dilution, especially if they arise from a small in volume gas component. Nevertheless, especially for low-$J$ CO transitions and oxygen lines it is reasonable to assume that the emission comes from a number of clouds spread accross the beam.

Table~\ref{tab:nclouds} shows our results. From this table we can see that for low-$J$ transitions the average densities for the individual clouds have to be $\lesssim 5000\,{\rm cm}^{-3}$ as above this density too few clouds of 2 mags in extinction are needed to match the observations. As regards the $4\le\,J<6$ CO transitions, these are coming from a higher density gas (but $\lesssim 10^5{\rm cm}^{-3}$), however very little material is needed in order to match these transitions. We also note that for $J > 6$ the absolute line intensities are always too high at $A_V\sim2\,{\rm mag}$.

We can see from the above analysis that the number of `clouds' necessary to match the different $J$ transitions of CO varies in a way consistent with the fact that different transitions will peak in different gas components. We note however that the chosen visual extinction of 2 mags is arbitrary: we can in fact determine the maximum Av that the individual clouds need to have in order for the model not to overestimate the observed emission. In Table \ref{tab:nclouds} we show additional results for $A_V=3$ and 5 mag. We can see that by 5 mag {\em all} the best fit models overstimate the absolute intensities of each transition, while $A_V=3\,{\rm mag}$ still gives reasonable solutions. We also find (not shown) that for $J > 6$ transitions, only models of clouds up to 1 mag in visual extinction lead to computed line intensities of individual lines less than or equal to the observed ones. In this latter case, we find that the best fits are for densities of $(1\sim3)\times10^5\,{\rm cm}^{-3}$ and a maximum number of clouds of $\sim1000$. Ultimately, however, we seek to find the {\it average} number of clouds emitting the observed radiation within our beam: from Table~\ref{tab:nclouds} and for $A_{\rm V,max}=2\,{\rm mag}$ we obtain $N_{\rm clouds}\sim10^4-10^5$.

Before discussing further a best fit model for all the CO transitions, it may be instuctive to analyse the behaviour of the individual observed line intensities shown in Fig.~\ref{fig:ratios1};  we find that for low-$J$ transitions we obtain a diagonal best-fit region from low densities and low UV strengths to high densities and high UV strengths. On the other hand, from Fig.~\ref{fig:ratios2} we find that for high-$J$ lines there is less dependence on the UV field strength and the best-fit regions are constrained to a narrower region with $n\sim10^{4.5}\,{\rm cm}^{-3}$ increasing to $\sim10^{5.2}\,{\rm cm}^{-3}$ or above for stronger UV fields. We argue that these higher $J$ transitions are mainly produced by regions where the density is high. This is in agreement with the fact that the critical density $n_{\rm crit}$ increases with $J_{\rm up}$ \citep{Oste74}. For example, at $T=100\,{\rm K}$, $n_{\rm crit}=1.8\times10^3\,{\rm cm}^{-3}$ for CO(1-0), $n_{\rm crit}=80.6\times10^3\,{\rm cm}^{-3}$ for CO(4-3), and $n_{\rm crit}=604\times10^3\,{\rm cm}^{-3}$ for CO(8-7).
At the temperatures we are considering, all CO lines with $J \ge 5$ have $n_{\rm crit} > 10^5\,{\rm cm}^{-3}$, which is consistent with the regions of best fit we find.

\begin{table*}
\caption{Number of model clouds required to reproduce the observed absolute line intensities for models with a maximum cloud extinction, $A_{V,{\rm max}}=2$~mag, based on the density limits, $n_\mathrm{H,lo}$ and $n_\mathrm{H,hi}$, derived from best-fit parameters for adjacent line ratios.}
\centering
\label{tab:nclouds}
\begin{tabular}{cc ccc ccc ccc ccc}
\hline
\hline
Line & Ratio &\multicolumn{3}{c}{$n_\mathrm{H,lo}$ (cm$^{-3}$)} & \multicolumn{3}{c}{$n_\mathrm{H,hi}$ (cm$^{-3}$)} & 
\multicolumn{3}{c}{$N_\mathrm{clouds}$ ($n_\mathrm{H,lo}$)} & \multicolumn{3}{c}{$N_\mathrm{clouds}$ ($n_\mathrm{H,hi}$)} \\
\multicolumn{2}{r}{}&2mag&3mag&5mag&2mag&3mag&5mag&2mag&3mag&5mag&2mag&3mag&5mag\\
\hline
CO(1--0) & (2--1)/(1--0) & 10$^{3.0}$      & 10$^{3.0}$      & 10$^{3.4}$  & 10$^{4.3}$      & 10$^{4.4}$ & 10$^{5.7}$  & 3$\times$10$^{5}$ & 1$\times$10$^{3}$ & 3.5  & 3       & $<$1  & 2$\times$10$^{-2}$  \\
CO(2--1) & (2--1)/(1--0) & 10$^{3.0}$      & 10$^{3.0}$      & 10$^{3.4}$  & 10$^{4.3}$      & 10$^{4.4}$ & 10$^{5.7}$  & 7$\times$10$^{4}$ & 4$\times$10$^{4}$ & 4.4 & $\sim$1 & 7  & 2$\times$10$^{-2}$  \\
CO(2--1) & (3--2)/(2--1) & 10$^{3.3}$      & 10$^{3.6}$      & 10$^{4.0}$  & 10$^{4.4}$      & 10$^{4.5}$ & 10$^{5.7}$  & 2$\times$10$^{5}$ & 1$\times$10$^{2}$ & 1$\times$10$^{-1}$ & $<$1    & $<$1  & 2$\times$10$^{-2}$  \\
CO(3--2) & (3--2)/(2--1) & 10$^{3.3}$      & 10$^{3.6}$      & 10$^{4.0}$  & 10$^{4.4}$      & 10$^{4.5}$ & 10$^{5.7}$  & 2$\times$10$^{5}$ & 6$\times$10$^{2}$ & 2$\times$10$^{-1}$  & $<$1    & $\sim$1  &  2$\times$10$^{-2}$ \\
CO(3--2) & (4--3)/(3--2) & 10$^{3.9}$      & 10$^{4.4}$      & 10$^{4.8}$  & 10$^{4.8}$      & 10$^{4.8}$ & 10$^{5.5}$  & 5$\times$10$^{3}$ & 5         & 4$\times$10$^{-2}$  & $\sim$1 & $\sim$1  & 2$\times$10$^{-2}$  \\
CO(4--3) & (4--3)/(3--2) & 10$^{3.9}$      & 10$^{4.4}$      & 10$^{4.8}$  & 10$^{4.8}$      & 10$^{4.8}$ & 10$^{5.5}$  & 4$\times$10$^{3}$ & 6         & 4$\times$10$^{-2}$ & $<$1    & $\sim$1  & 2$\times$10$^{-2}$  \\
CO(4--3) & (5--4)/(4--3) & $\le$10$^{2.0}$ & $\le$10$^{2.0}$ & 10$^{4.6}$  & 10$^{4.6}$      & 10$^{4.5}$ & 10$^{4.8}$  & 8$\times$10$^{6}$ & 8$\times$10$^{6}$ & 1$\times$10$^{-1}$  & $\ll$1  & $\sim$1  & 9$\times$10$^{-2}$  \\
CO(5--4) & (5--4)/(4--3) & $\le$10$^{2.0}$ & $\le$10$^{2.0}$ & 10$^{4.6}$  & 10$^{4.6}$      & 10$^{4.5}$ & 10$^{4.8}$  & 8$\times$10$^{6}$ & 7$\times$10$^{6}$ & 2$\times$10$^{-1}$  & $\ll$1  & $\sim$1  & 9$\times$10$^{-2}$  \\
CO(5--4) & (6--5)/(5--4) & 10$^{4.2}$      & 10$^{5.0}$      & 10$^{5.4}$  & 10$^{5.4}$      & 10$^{5.8}$ & 10$^{5.8}$  & 2$\times$10$^{3}$ & $<$1       & 1$\times$10$^{-2}$  & $\sim$1 & $<$1  & 5$\times$10$^{-3}$  \\
CO(6--5) & (6--5)/(5--4) & 10$^{4.2}$      & 10$^{5.0}$      & 10$^{5.4}$  & 10$^{5.4}$      & 10$^{5.8}$ & 10$^{5.8}$  & 3$\times$10$^{3}$ & $\sim$1    & 2$\times$10$^{-2}$  & $\sim$1 & $<$1  & 5$\times$10$^{-3}$  \\
CO(6--5) & (7--6)/(6--5) & $\le$10$^{2.0}$ & $\le$10$^{2.0}$ & 10$^{5.1}$  & 10$^{5.2}$      & 10$^{5.6}$ & 10$^{5.7}$  & $\gg$10$^{6}$     & $\gg$10$^{6}$ & 7$\times$10$^{-2}$  & $\sim$1 & $\ll$1  &8$\times$10$^{-3}$  \\
CO(7--6) & (7--6)/(6--5) & $\le$10$^{2.0}$ & $\le$10$^{2.0}$ & 10$^{5.1}$  & 10$^{5.2}$      & 10$^{5.6}$ & 10$^{5.7}$  & $\gg$10$^{6}$     & $\gg$10$^{6}$ & 7$\times$10$^{-2}$  & $\sim$1 & $\ll$1  &7$\times$10$^{-3}$  \\
CO(7--6) & (8--7)/(7--6) & 10$^{5.4}$      & -n/a-           & -n/a-       & $\ge$10$^{6.0}$ & -n/a-      & -n/a-       & $\sim$1           & -n/a-     & -n/a-  & $\ll$1  & -n/a- & -n/a- \\
CO(8--7) & (8--7)/(7--6) & 10$^{5.4}$      & -n/a-           & -n/a-       & $\ge$10$^{6.0}$ & -n/a-      & -n/a-       & -n/a-             & -n/a-     & -n/a-  & -n/a-   & -n/a- & -n/a- \\
\hline
\end{tabular}
\end{table*}

\subsection{Best-fit UV field strengths based on PDR fine structure line diagnostics}
\label{ssec:UVestimate}

\begin{figure}
\begin{center}
\includegraphics[width=9cm]{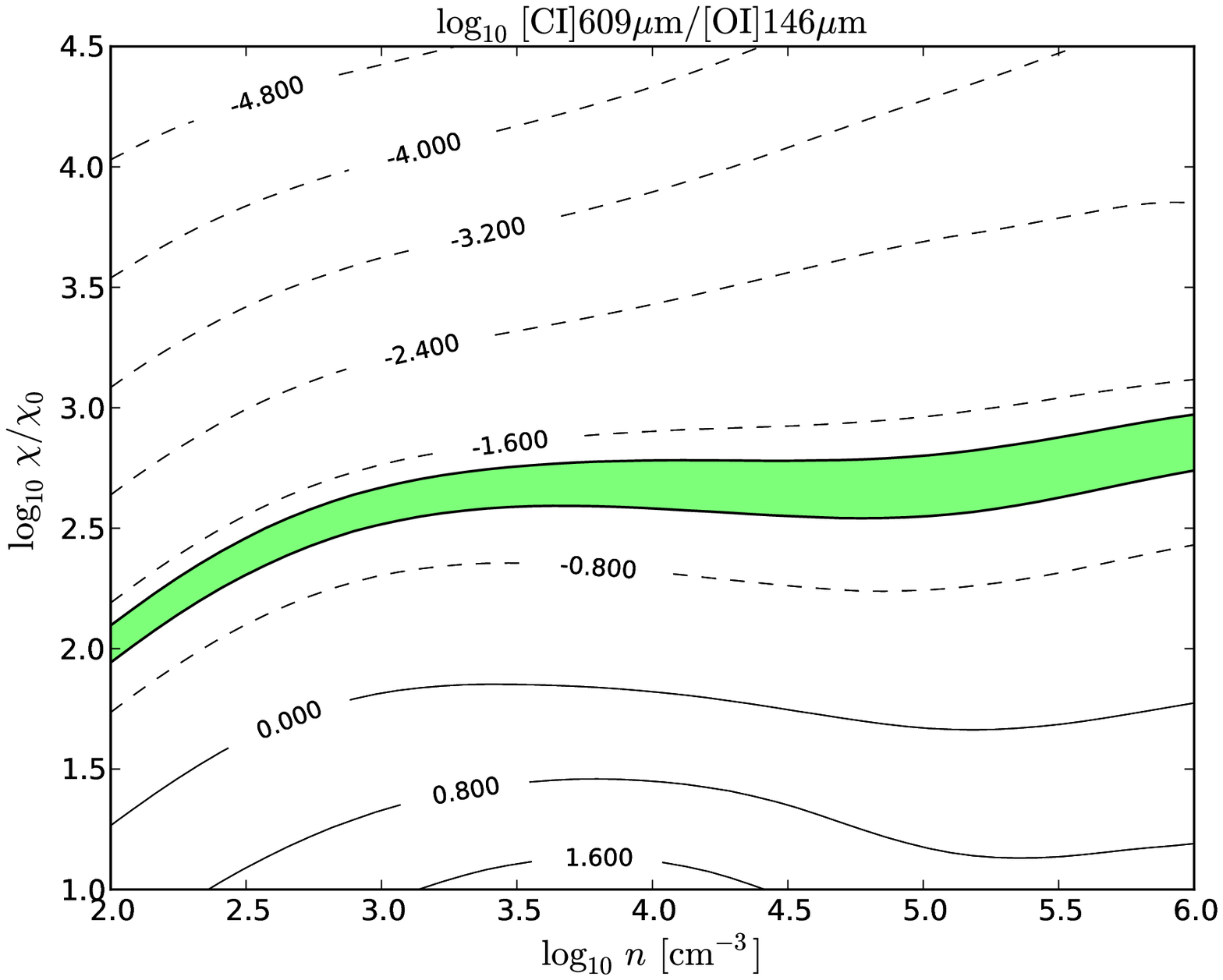} 
\includegraphics[width=9cm]{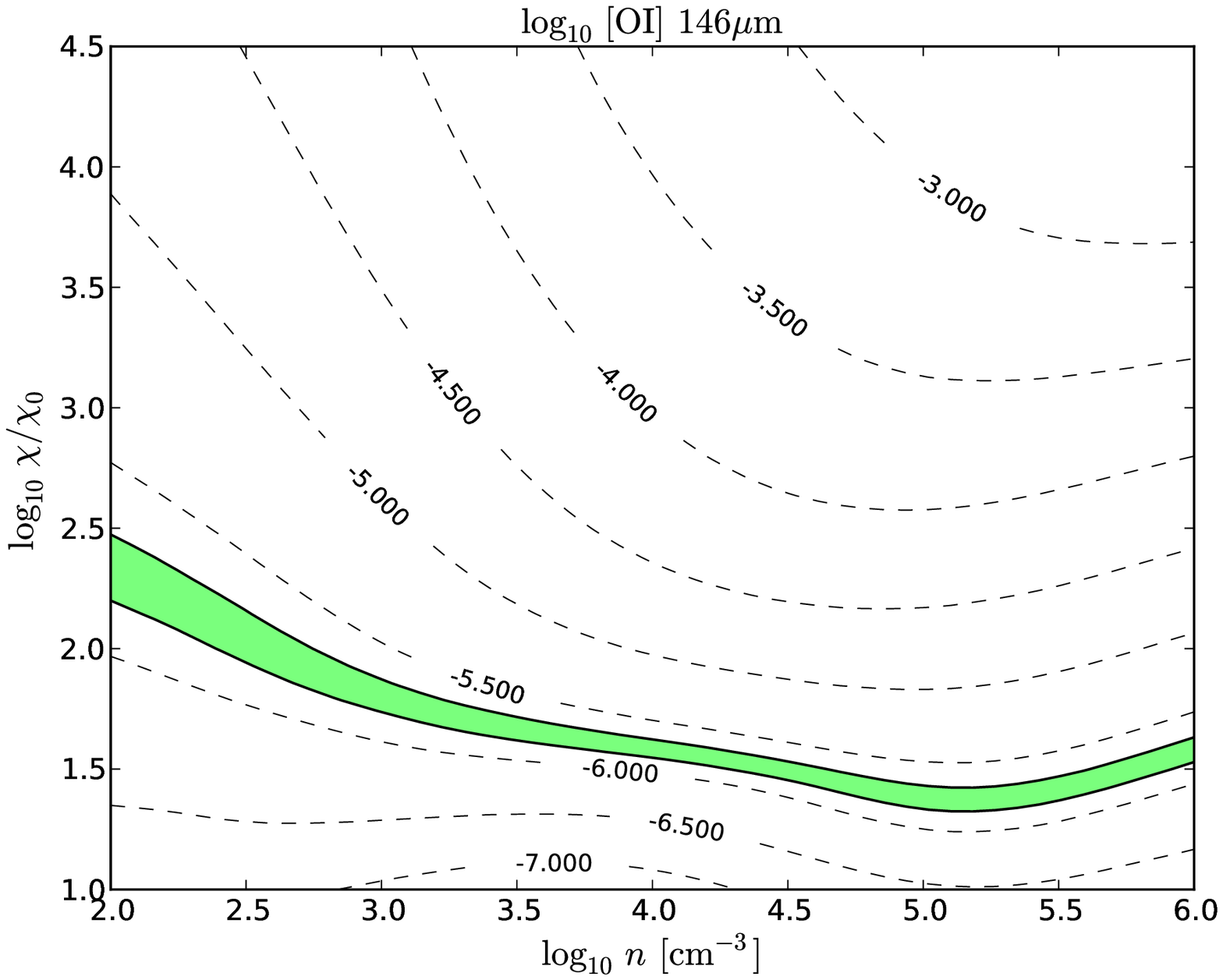} 
\caption{Top panel: CSO and {\it ISO}-LWS observations of the {\ci/\oib} ratio for NGC 4038. We find that the strength of the UV field is in general $\chi\sim10^{2.5}\chi_{0}$. Bottom panel: {\it ISO}-LWS observations of the {\oib} line. This line suffers from beam dilution and we find that it is highly underestimaed by a factor of $\sim10^{-2}$.}
\label{fig:flines}
\end{center}
\end{figure}

The fine structure lines of [C\,{\sc ii}], [C\,{\sc i}] and [O\,{\sc i}] are frequently used as diagnostics for the PDR properties \citep[e.g. ][]{Kauf99}. In the top panel of Fig.~\ref{fig:flines}, we use the {\ci/\oib} line ratio as a diagnostic for the UV field. Although, as we have mentioned above, the local emissivity of the {\ci} line peaks at intermediate depths, the best-fit models for this ratio indicate the presence of a strong radiation field, which is required to heat the gas in order to excite the {\oib} line. We find here that the UV field is in general $\chi\gtrsim10^{2.5}\chi_{0}$. We note that, of the fine structure lines, only {\ci} was observed using a different telescope (CSO) and thus is subject to a different amount of beam dilution. If the {\ci} emission is more diluted than we have assumed when convolving to the 80\arcsec\ beam, then the best-fit region in the middle panel of Fig.\ref{fig:flines} would move upward, implying stronger best-fit UV field strengths.

For the {\oib} line intensity, which is shown in the bottom panel of Fig. \ref{fig:flines}, the model line emission shows little sensitivity to density and is instead constrained by the UV field. We find that the observed line intensity is in general reproduced by models with a UV field strength of $\chi\sim10^{1.5}$--$10^{2.5}\chi_{0}$. If the {\oib} emission is diluted by the beam, then this would suggest that a stronger UV field would be required. This in turn would lead to better agreement with the UV field implied by the {\ci}/{\oib} ratio. We find that in order to reproduce the UV field strength implied by the {\cii}/{\oib} ratio, the required beam dilution factor would be $\sim10^{-2}$.

We note that the {\cii} emission is matched by a very low radiation field, lower than $10\chi_{0}$, which contradicts the outcomes we find from all the other lines. We explain this by the following scenario: if the PDR surfaces cover the region and fill the beam, then the lower value cannot be a result of beam dilution. Therefore, either the PDRs are not uniformly distributed across the region, or their outermost layers are very thin, significantly less than $A_V=2\,{\rm mag}$. For instance, for $n=100\,{\rm cm}^{-3}$ and $\chi=10^3\chi_{0}$, we reproduce the observed {\cii} intensity with a PDR layer of only $A_V\simeq0.2\,{\rm mag}$. For model clouds of the same density interacting with $\chi=10^{3.5}\chi_{0}$, we match the observed intensity for $A_V\simeq0.25\,{\rm mag}$, while for $n=10^3\,{\rm cm}^{-3}$ and $\chi=10^{3.5}\chi_{0}$, only $A_V\simeq0.04\,{\rm mag}$ of PDR material is needed to reproduce the line intensity. In all cases $A_V<2\,{\rm mag}$. 

In addition, as discussed above, the {\oia} line suffers from self-absorption, the amount of which is not constant. \citet{Vast10} used the {\sc smmol} radiative transfer code \citep{Rawl01} to model the [O\,{\sc i}] emission from PDRs and found that the {\oia} line intensity may be reduced by 20--80\%, depending on the PDR parameters. We have adopted the assumption that the clouds have distinct velocities along the line of sight. Under this assumption the intensity is not prone to mutual shielding and self-absorption and therefore the clouds do not attenuate.

In this paper we choose not to present ratios involving fine structure lines with CO lines, as these lines have been observed by different instruments and thus the beam dilution factor is not known. This is of crucial importance as the {\cii} fine structure line is emitted from the outermost and thinnest parts of PDRs, contrary to the CO lines which are emitted from a wider region in the innermost parts of PDRs, as discussed in the introduction to this section.

\subsection{Unified picture}
\label{ssec:unified}

\begin{figure*}
\begin{center}
\includegraphics[width=0.9\textwidth]{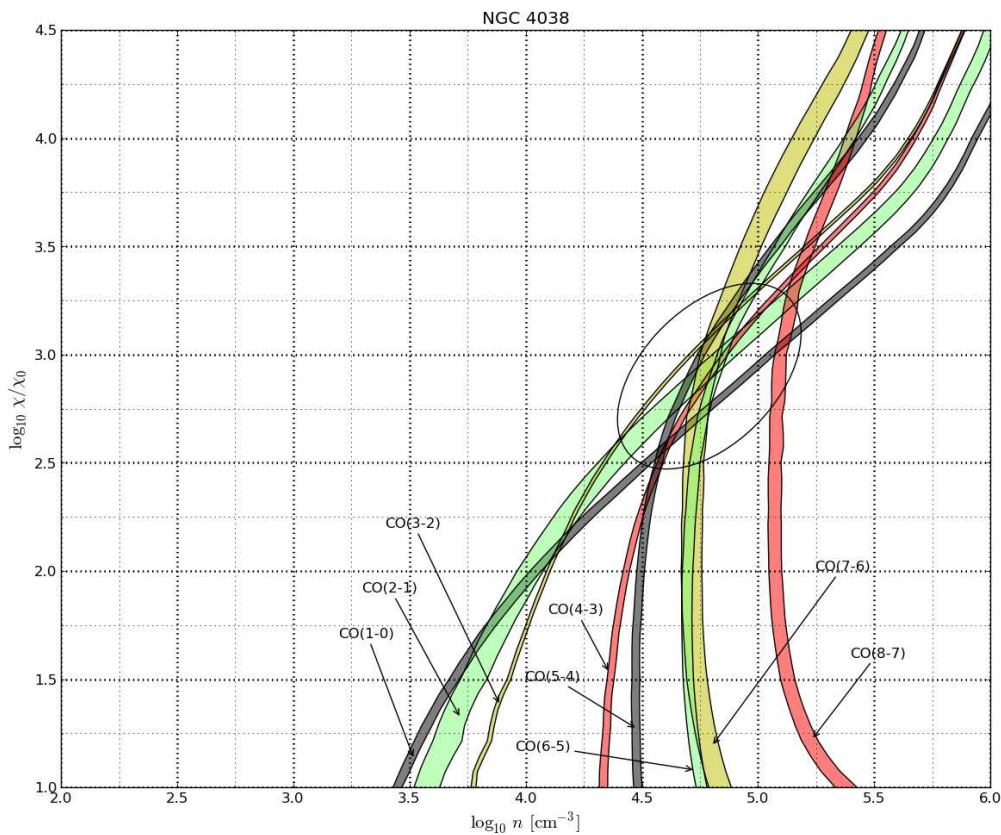}
\caption{This figure summarises the agreement of our models with observations for the CO $J=(1-0)..(8-7)$ line intensities. All lines have been convolved to the {\it ISO}-LWS 43\arcsec beam size. The particular error bars of CO(4-3) and CO(7-6) lines are to include the observational values of both instruments (see ID=4a,4b and 7a,7b of Table \ref{tab:observ}). We find an $n-$UV best-fit region in which the line intensities are converging corresponding to PDRs of high densities interacting with high UV field strengths; i.e. $n\sim10^{4.5}-10^{5.2}\,{\rm cm}^{-3}$ interacting with $\chi\sim10^3$--$10^4\chi_{0}$. Note that there are no best-fit regions for $n<10^3\,{\rm cm}^{-3}$ for any CO transition line.}
\label{fig:all}
\end{center}
\end{figure*}

From our analysis so far we see that different transitions are tracing gas components characterised by different excitations and densities. We thus argue that within the low spatial resolution beams available to extragalactic studies, it is reasonable to exclude discussions about {\it average} conditions, i.e. densities or radiation fields. Nevertheless, it can be instructive to plot all our results together and draw a summary of our findings. From Fig.~\ref{fig:all}, where we plot the best fits for all the observed CO line intensities in a density-UV plane, we find the following quantitative results:
\begin{enumerate}[i)]
\item The CO(1--0) transition can be fitted by the largest range of $n$--UV combinations, where a low density requires a low radiation field strength and vice-versa. Taken in isolation this transition does not provide us with any information on the density or radiation field of the gas.
\item As we move towards higher transitions, the parameter space is more constrained in density, i.e the CO(6--5) and CO(7--6) lines are best fitted by densities between $10^{4.5}-10^{5.3}\,{\rm cm}^{-3}$.
\item Although the best fit for each transition is given by a different model, there is a $n$--UV space convergence in the ranges of $10^4$--$10^{5.3}\,{\rm cm}^{-3}$ in density and $\chi\sim10^{2.5}$--$10^4\chi_{0}$ in radiation strength.
\end{enumerate}

We see that all CO lines converge towards higher best-fit densities and UV field strengths than those implied by the fine structure lines. A possible explanation for this is that some fraction of the CO line emission arises from gas heated by other mechanisms, such as turbulent heating, supernova-driven shocks or shocks produced by colliding giant molecular clouds. However, we do not perform any further analysis here to study these different mechanisms.

In summary, while it may make sense to derive an average radiation field for the region (and hence obtain an idea of energetics), it is less meaningful to infer a single best fit value for the density of the gas.

\subsection{Comparison with other models}
\label{ssec:comparison}

Our findings from the unified picture are generally in agreement with those of other authors who have attempted to model the Antennae galaxies, pointing to a stronger-than-Galactic UV field together with a gas component at high enough density to reproduce the observed intensities. Early work by \citet{Fisc96} using {\it ISO}-LWS observations found that the typical cloud properties of PDRs corresponded to densities of $n\sim(0.25-1.5)\times10^4\,{\rm cm}^{-3}$ interacting with UV fields of $200-2500$ times the local interstellar radiation field (ISRF), and that their temperatures were about $200\pm60\,{\rm K}$. \citet{Niko98} used {\cii} observations and assumed a single emission component in the beam to infer that the average density in NGC 4038 is $\sim10^5\,{\rm cm}^{-3}$ with a UV radiation field of $\sim500\chi_{0}$. They also found that the bulk of the {\cii} emission comes from PDRs and that only a minor fraction of the line emission can come from ionised gas in H\,{\sc ii} regions. \citet{Gilb00} also found that the observed emission lines of NGC 4038/9 can be explained using a best fit model with density $n\sim10^4-10^5\,{\rm cm}^{-3}$ and UV field of $\sim5\times10^3\chi_{0}$. In contrast, \citet{Baye06} derived a best fit density $n \sim 3.5\times10^5\,{\rm cm}^{-3}$ and a UV field strength $\chi \sim 10^{5.4}\chi_{0}$. These latter values, especially those of \citet{Baye06}, are significantly higher than those we infer. \citet{Schu07} analysed CO $J=1$--0 to $J=3$--2 line observations and argued that the average ISRF is between 500 and $3000\chi_{0}$, and that most of the CO emission arises from small and moderately dense clumps with $n \lesssim 5\times10^4\,{\rm cm}^{-3}$. These authors used low-$J$ transition lines for their analysis which corresponds to the first crossing shown in Fig.\ref{fig:all}. However including higher $J$ transition lines shifts the best-fit area to higher densities interacting with stronger UV radiation fields. Hence such additional information may give rise to a different interpretation of the data.

In the work by S13, the PDR models they used suggested the presence of a warm gas component with an average ISRF of $\chi\sim10^3\chi_{0}$ and a cold component exposed to $\chi\sim10^2\chi_{0}$. They also claim that these values are consistent with all three regions of the Antennae (the NGC 4038/9 nuclei and the overlap region). For their PDR models they used the results of \citet{Holl12}. To make a more quantitative comparison, we have also run a grid of simulations using the same metallicity as they used, corresponding to standard Milky Way abundances. 
We found that by increasing the abundances above the Milky Way values, they act to lower the UV field needed to reproduce the observed line intensities. This is consistent with the fact that greater abundances compensate for the lower temperatures produced by weaker UV fields. 


\subsection{X-factor}
\label{ssec:xfactor}

Due to its lack of a dipole moment, molecular hydrogen is not directly detectable at the low temperatures found in the bulk of the ISM, and alternative methods have therefore been developed to trace it. The most common is through the so-called $X_{\rm CO}$ factor, in which the CO emission is related to the column density of molecular hydrogen, ${\rm N}({\rm H}_2)$, along a line of site, although early work by \citet{Papa04, Bell07} and recently by \citet{Offn14} has found that in some instances atomic carbon might be a better tracer of molecular hydrogen. The theoretical definition of the $X_{\rm CO}$ is given by:
\begin{eqnarray}
\label{eqn:xfactor}
X_{\rm CO}=\frac{{\rm N}({\rm H}_2)}{\int T_{\rm A}({\rm CO})dr}\,\left[{\rm cm}^{-2}\,({\rm K}\,{\rm km}\,{\rm s}^{-1})^{-1}\right]\,,
\end{eqnarray}
where the denominator corresponds to the integrated antenna temperature of ${\rm CO}(1-0)$ line. 

\begin{figure}
\begin{center}
\includegraphics[width=9cm]{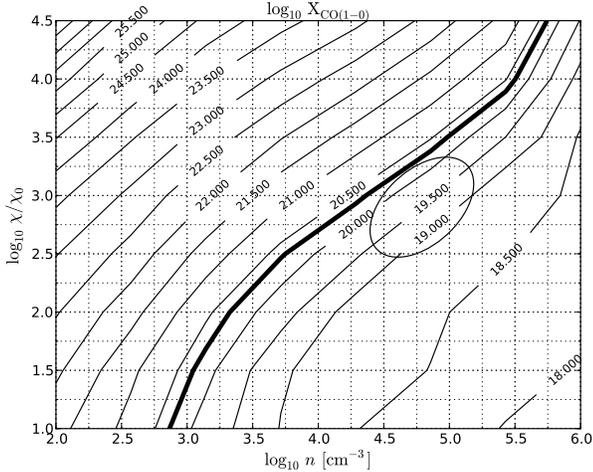} 
\caption{This figure shows contours of the $X_{{\rm CO}(1-0)}$ factor as a function of density and UV field strength. The thick solid line corresponds to the canonical value of the Milky Way of $\sim10^{20.3}\,{\rm cm}^{-2}\,\left({\rm K}\,{\rm km}\,{\rm s}^{-1}\right)^{-1}$. The ellipse corresponds to the best-fit area as obtained from Fig. \ref{fig:all}. We argue that these are the best values of the $X_{{\rm CO}(1-0)}$ factor to estimate the mass of molecular Hydrogen, ${\rm H}_2$, in the different regions of NGC 4038.}
\label{fig:xfactor}
\end{center}
\end{figure}

From Fig.\ref{fig:all} we find that the best fit models must satisfy a relation between the density and the strength of the UV radiation field. As discussed earlier, low density regions may in principle interact with strong UV fields and higher density regions with weaker UV fields. Given that the column density of molecular hydrogen and the integrated intensity of the CO(1--0) line are computed by the PDR model along the line of sight into the cloud, we are able to produce a map of $X_{\rm CO}$ factor values as a function of parameter space for our grid of simulations. This map is shown in Fig.~\ref{fig:xfactor}. The contours correspond to $\log X_{\rm CO}$ (see Eqn.\ref{eqn:xfactor}). Comparing the best-fit regions indicated by Fig.\ref{fig:all}, we find a range of values for $X_{\rm CO}$. For example, in the $n$--UV region where the CO lines converge (i.e. $n \sim 10^{4.8}\,{\rm cm}^{-3}$, $\chi \sim 10^{2.9}\chi_{0}$), we find $X_{\rm CO} \lesssim 10^{20}\,{\rm cm}^{-2}\,({\rm K}\,{\rm km}\,{\rm s}^{-1})^{-1}$, which is less than the canonical Milky Way value of $\sim$$2 \times 10^{20}\,{\rm cm}^{-2}\,({\rm K}\,{\rm km}\,{\rm s}^{-1})^{-1}$ \citep[][]{Stro96, Dame01}. The general trend in the $X_{\rm CO}$ values shown in Fig.~\ref{fig:xfactor} is consistent with that found by \citet{Bell06}. The X-factor decreases as the density rises, and increases as the UV field becomes stronger. Our finding that the $X_{\rm CO}$ factor of NGC4038 is less than the canonical Milky Way value is in agreement with the $X_{\rm CO}$ factors derived for other AGN and starburst galaxies \citep{Bell07}.

\section{Discussion}
\label{sec:disc}


For the simulations presented in this paper, the mass of a spherical cloud corresponding to a given one-dimensional slab model is determined by the equation:
\begin{eqnarray}
\label{eqn:mass}
M=\frac{4\pi}{3}R^3n_{_{\rm H}}m_{_{\rm H}}\mu
\end{eqnarray}
where $R$ is the maximum depth reached in the PDR model, $n_{\rm H}$ is its H-nucleus density, $m_{\rm H}$ is the Hydrogen mass, and $\mu=1.36$ is a mass correction factor to account for the contribution from Helium and heavier elements.

Given the cloud density ranges inferred from the CO line ratios and the associated number of clouds required to reproduce the observed line intensities (see \S\ref{ssec:nclouds}), we can derive estimates for the total mass of the molecular gas contained in these model clouds. Adopting our maximum $A_V$ of 2 mag and the $n_\mathrm{H}$ and $N_\mathrm{cloud}$ values from Table \ref{tab:nclouds}, we find masses for individual clouds that vary from $<$10 ${\rm M}_\odot$ up to $\sim$10$^{4}$ ${\rm M}_\odot$, and total masses from all clouds of 10$^{3}$ to 10$^{10}$ $M_\odot$. While we can immediately rule out some of the inferred model scenarios as being unphysical (e.g., those leading to individual cloud masses less than 10 $M_\odot$), the range of total masses may form two extremes of the ISM conditions. The low total mass estimates could indicate that the emission we are reproducing with our models comes from the surface layers of numerous, larger clouds, in which case these masses are those of the total gas contained in PDRs spread across the telescope beam. Alternatively, the upper end of the total mass range is comparable to estimates of the total virial mass within NGC 4038 \citep[e.g. ][]{Wils00}, in which case our model clouds would represent the entirety of the molecular gas contained within this region, and consequently imply that {\em all} molecular gas resides within smaller, low $A_V$ clouds that fill the beam. For the best-fit region of parameter space described in \S\ref{ssec:unified} we find that the observed gas corresponds to a total mass of $\sim3\times10^4\,{\rm M}_{\odot}$. This mass corresponds to the warmer PDR component only, and not to the total gas mass, which includes the cold component (at $\sim10\,{\rm K}$) deep within GMCs.

We note that the larger total mass estimates come from the lower end of the density range we infer from the low-$J$ CO line ratios, indicating that, under this scenario, the molecular gas is present within more typical GMCs of larger size and intermediate density. However, the higher densities that we infer from the high-$J$ CO line ratios are more consistent with those implied by the fine structure lines, which would mean that the emission we are modelling is that of the UV-illuminated surfaces of larger cloud structures.

Regarding the adopted distance of $22\pm3\,{\rm Mpc}$ \citep{Schw08} described in the Introduction of the present work, we note that \citet{Savi08} have estimated a much smaller distance to the Antennae of about $13\,{\rm Mpc}$. If we assumed the latter distance, we note that the results described in this paper would not be altered, as our basic model representation does not depend on the distance to NGC 4038 and thus would remain unaffected. This applies also to our derived cloud properties for the best-fit range of densities and UV field strangths. However, assuming a smaller distance would affect the determination of the molecular mass from the $X_{\rm CO}$ factor when using the total CO(1-0) luminosity instead its integrated intensity.


\section{Conclusions}\label{sec:conc}

In this paper we use the {\sc 3d-pdr} code to produce a grid of $\sim1400$ PDR models in order to understand the conditions of the ISM occuring in the NGC 4038 nucleus of the Antennae interacting galaxies complex, using observations of various lines taken from ground-based telescopes such as SEST and CSO, as well as from space telescopes such as {\it Herschel} and {\it ISO}.

The PDR clouds are considered to be one-dimensional slabs with uniform densities in the range $10^2<n_{_{\rm H}}<10^6\,{\rm cm}^{-3}$, interacting with various plane-parallel interstellar radiation fields in the range $10<{\chi}/{\chi_{0}}<10^{4.5}$. We have integrated all line intensities up to a cloud depth of $A_V=2\,{\rm mag}$. Integrating to larger depths leads to absolute line intensities that are greater than the observed ones. We adopted solar abundances and used a subset of the most recent UMIST data base (UMIST 2012) of rates and reactions. We analysed the model cloud emission in the {\ci} and {\cii} fine structure lines and eight CO transition lines with $J=1-0$ up to $J=8-7$. 

We find that the number of cloud surfaces needed to match the different CO $J$ transitions varies, as different transitions will be tracing different gas components. For low-$J$ transitions we obtain a diagonal best-fit region extending from low densities and low UV field strengths to high densities and high UV field strengths. For high-$J$ transitions we find a weak dependence on the UV field and the best-fit region constrains the densities to be $n\sim10^{4.5}-10^{5.2}\,{\rm cm}^{-3}$. Since a fairly wide density range is needed to explain the CO line intensities, we argue that using a single component model to explain all results is not meaningful, as it leads to large uncertainties in the derived properties. The {\ci}/{\oib} ratio shows that the UV field is in general $\chi\gtrsim10^{2.5}\chi_{0}$ and a beam dilution factor of the order of $\sim10^{-2}$ is needed to reproduce the observed line emission from the modelled cloud.

Furthermore, we find that the the total mass in all clouds is estimated to be 1000 ${\rm M}_{\odot}$ up to $10^{10}\,{\rm M}_{\odot}$. This range of total masses may correspond to two extremes of the ISM conditions: the low total mass end would indicate emission from the surfaces of large clouds that suffer from beam dilution, while the high total mass end indicates that all molecular gas resides within smaller clouds that fill the beam. The best-fit parameters obtained by fitting the CO line intensities suggest that the mass of gas in PDRs corresponds to $\sim3\times10^4\,{\rm M}_{\odot}$.

Finally, we find that the CO-to-H$_2$ X-factor value is less than the canonical Milky Way value, in agreement with the values found for AGN and other starburst galaxies \citep{Isra06,Isra09a,Isra09b,Bell07,Bola13}.

Our future work will include fully three-dimensional calculations for the Antennae galaxies system using Smoothed Particle Hydrodynamics density distributions \citep{Karl10,Karl13} and the newly implemented code \texttt{TORUS-3DPDR} (Bisbas et al. {\it in prep.}) which includes a full three-dimensional calculations of photoionization.

\section*{Acknowledgements}
We thank the anonymous referee whose comments and suggestions have significantly improved the clarity of the paper. The work of TGB was funded by STFC grant ST/J001511/1. TGB acknowledges the NORDITA program on Photo-Evaporation in Astrophysical Systems (June 2013) where part of the work for this paper was carried out. TAB thanks the Spanish MINECO for funding support from grants CSD2009-00038, AYA2009-07304, and AYA2012-32032. TAB is supported by a CSIC JAE-DOC research contract. This research has made use of NASA's Astrophysics Data System.

This work used the DiRAC Data Analytic system at the University of Cambridge, operated by the University of Cambridge High Performance Computing Service on behalf of the STFC DiRAC HPC Facility (www.dirac.ac.uk). This equipment was funded by BIS National E-infrastructure capital grant (ST/K001590/1), STFC capital grants ST/H008861/1 and ST/H00887X/1, and STFC DiRAC Operations grant ST/K00333X/1. DiRAC is part of the National E-Infrastructure.



\end{document}